\documentclass[]{aa} 
\usepackage{amsmath}
\usepackage{txfonts} 
\usepackage{graphicx}
\usepackage{subfigure}
\usepackage{color}
\usepackage{bm}
\usepackage{natbib}
\usepackage{url}
\def\ltsima{$\; \buildrel < \over \sim \;$}
\newcommand{\simgt}%
        {\,\hbox{\lower0.6ex\hbox{$\sim$}\llap{\raise0.6ex\hbox{$>$}}}\,}
\def\simlt{\lower.5ex\hbox{\ltsima}}
\begin{document} 
\title{Toward understanding the formation of multiple systems\thanks{Based on observations carried out with the IRAM Plateau de Bure Interferometer. 
IRAM is supported by INSU/CNRS (France), MPG (Germany), and IGN (Spain).}}
\subtitle{A pilot IRAM-PdBI survey of Class 0 objects} 
\author{A. J. Maury\inst{1} 
\and Ph. Andr\'e\inst{1} 
\and P. Hennebelle\inst{2}
\and F. Motte\inst{1}
\and D. Stamatellos\inst{3}
\and M. Bate\inst{4}
\and A. Belloche\inst{5}
\and G. Duch\^ene\inst{6,7}
\and A. Whitworth\inst{3}} 
\institute{Laboratoire AIM, CEA/DSM-CNRS-Universit\'e Paris Diderot, IRFU/Service d'Astrophysique, C.E. Saclay, 
Orme des Merisiers, 91191 Gif-sur-Yvette, France
\and Laboratoire de radioastronomie, UMR 8112 du CNRS, Ecole normale sup\'erieure et Observatoire de Paris, 24 rue Lhomond, 75231 Paris, France 
\and School of Physics \& Astronomy, Cardiff University, Cardiff, CF24 3AA, Wales, UK
\and School of Physics, University of Exeter, Stocker Road, Exeter EX4 4QL, UK
\and Max-Planck Institut f\"ur Radioastronomie, Auf dem H\"ugel 69, 53121 Bonn, Germany
\and Astronomy Department, University of California, Berkeley, CA 94720-3411, USA
\and Laboratoire d'Astrophysique de Grenoble, Universit\'e Joseph Fourier, BP 53, 38041 Grenoble cedex 9, France} 
\date{Received 20 October 2009 / Accepted 13 January 2010} 
\abstract 
{The formation process of binary stars and multiple systems is poorly understood. The multiplicity rate of Class~II pre-main-sequence stars and Class~I protostars is well documented and known to be high ($\sim$ 30\% to 50\% between $\sim$100 and 4000~AU). However, optical / near-infrared observations of Class~I/Class~II YSOs barely constrain the pristine properties of multiple systems, since dynamical evolution can quickly alter these properties during the protostellar phase.} 
{Here, we seek to determine the typical outcome of protostellar collapse and to constrain models of binary formation by core fragmentation during collapse, using high-resolution millimeter continuum imaging of very young (Class~0) protostars observed at the beginning of the main accretion phase.} 
{We carried out a pilot high-resolution study of 5 Class 0 objects, including 3 Taurus sources and 2 Perseus sources, using the most extended (A) configuration of the IRAM Plateau de Bure Interferometer (PdBI) at 1.3~mm. Our PdBI observations have a typical HPBW resolution $\sim$0.3 \arcsec -- 0.5 \arcsec ~and rms continuum sensitivity $\sim$ 0.1 -- 1~mJy/beam, which allow us to probe the multiplicity of Class~0 protostars down to separations $a \sim$50~AU and circumstellar mass ratios $q \sim$0.07.} 
{We detected all 5 primary Class~0 sources in the 1.3~mm dust continuum. A single component associated 
with the primary Class 0 object was detected in the case of the three Taurus sources, while robust 
evidence of secondary components was found toward the two Perseus sources: L1448-C and NGC1333-IR2A. 
We show that the secondary 1.3 mm continuum component detected $\sim$ 600~AU south-east of L1448-C, at a position angle close to that of the CO(2--1) jet axis traced by our data, is an outflow feature directly associated with the powerful jet driven by L1448-C. The secondary 1.3 mm continuum component detected $\sim$ 1900~AU south-east of NGC1333-IR2A may either be a genuine protostellar companion or trace the edge of an outflow cavity.
Therefore, our PdBI observations revealed only wide ($>$ 1500~AU) protobinary systems and/or outflow-generated features.}
{When combined with previous millimeter interferometric observations of Class~0 protostars, our pilot PdBI study tentatively suggests that the binary fraction in the $\sim$ 75 -- 1000 AU range increases from the Class 0 to the Class I stage. It also seems to argue against purely hydrodynamic models of binary star formation. We briefly discuss possible alternative scenarios to reconcile the low multiplicity rate of Class~0 protostars on small scales with the higher binary fraction observed at later (e.g. Class~I) evolutionary stages.}
\maketitle 

\section{Introduction} 

Understanding the formation of multiple systems is a major unsolved problem in 
star formation research (e.g. \citealt{Tohline02}). 
While most solar-type (0.5 $\leq$ M$_{\odot} \leq$ 2 M$_{\odot}$) pre-main sequence (PMS) stars are observed to be in binary  
systems with typical separations $\sim $~10--300~AU (e.g., \citealt{Duchene04, Duchene07}, see also below), 
the detailed manner in which individual prestellar cores fragment (or not) during collapse to form multiple (or single) stars is still poorly understood and highly debated (see \citealt{Goodwin07} for a review). Even the typical outcome of cloud core collapse is unclear since it has been argued that most stars may actually form as single objects \citep{Lada06}. 
The argument is based on the fact that most {\it low-mass} stars (with $M_\star < 0.5\, M_\odot $) are single and that the stellar initial mass function (IMF) 
is significantly more populated below $0.5\, M_\odot $ than above (e.g. \citealt{Kroupa01,Chabrier05}). 
Despite conventional wisdom, it is therefore conceivable that most low-mass prestellar cores may collapse to single stars.\\
It is generally believed that multiple systems form by dynamical rotationally-driven fragmentation
at the end of (or shortly after) the first collapse phase of prestellar cores when the central $H_2$ density reaches $n_{crit} \sim 3 \times 10^{10}\, {\rm cm}^{-3}$ and the equation of state 
of the gas switches from isothermality to adiabacity \citep{Goodwin07}. 
Purely hydrodynamic SPH simulations of rotating cloud core collapse show that a very low level of initial core turbulence (e.g. $E_{turb}/E_{grav} \sim $~5\%)  leads to the formation of a multiple system \citep{Goodwin04, Hennebelle04, Commercon08}. 
In such SPH simulations, fragmentation is driven by a combination of rotation/turbulence and occurs in large ($\simgt 100$~AU) disk-like structures or "circumstellar accretion regions'' (CARs -- cf. \citealt{Goodwin07}). 
These CARs are not rotationally supported and are highly susceptible to spiral instabilities which fragment them into small-$N$ {\it multiple systems  with $N > 2$}, typically $N \sim $~3-4 components at radii $\leq 150$~AU in the equatorial plane \citep{Goodwin04, Fromang06}.
\\However, a very different outcome is found in simulations of {\it magnetized}
core collapse, as shown by recent results obtained with MHD codes
using both grid techniques \citep{Fromang06, Hennebelle08b, Machida05, Mellon09}
and the SPH technique \citep{Price07}. These new MHD simulations indicate 
that the presence of an even moderate magnetic field strongly modifies 
angular momentum transport during collapse and at least partly suppresses core 
fragmentation, often leading to the formation of a {\it single} object.  
\citet{Price07} and \citet{Hennebelle08b} conclude that binary star formation is 
still possible in the presence of magnetic fields but either requires strong initial perturbations 
or must occur during the second collapse phase, after the dissociation of $H_2$ \citep{Machida08}. 
The systems formed in the latter case are initially very-low-mass ($\sim 0.01\, M_\odot $), close 
($\sim 1 $~AU) binaries \citep{Bonnell94b}, which have to grow substantially by 
accretion during the Class~0/Class~I phase \citep{Bate00a} to match the properties of 
observed young binary stars \citep[e.g.][] {Duchene04}. 
Therefore, both from an observational and a theoretical point of view, it is unclear whether the collapse of 
a prestellar core typically produces one, two, or more stars. 

The multiplicity of solar-type pre-main-sequence stars and Class~II/Class~I  young stellar objects (YSOs) is now well
documented and has been investigated at a range of wavelengths  
(e.g., \citealt{Ghez93, Simon95, Patience02, Duchene04, Duchene07, Connelley08b}).
In particular, \citet{Patience02} and \citet{Kohler08} showed that the binary frequency of (Class~II) T Tauri stars ranges from $\sim$40\% to $\sim$60\% with a peak in the separation distribution around 60$^{+40}_{-20}$~AU.
\citet{Duchene07} showed that $\sim$32\% of Class~I YSOs have companions in the 
$\sim$50--$1000$ AU separation range. The observed Class~I/II binary systems have 
typical mass ratios $q = M_{\rm{second}} / M_{\rm{main}}\sim$ 0.2--1 (e.g. \citealt{Woitas01}). 
Unfortunately, observations of Class~II/Class~I YSOs barely constrain the {\it pristine 
properties of multiple systems} since dynamical evolution can drastically alter these properties
in less than $\sim 10^5$~yr \citep[cf.][]{Reipurth01b}. 
Although still fairly uncertain and a matter of debate, the lifetime of the Class~0 phase is estimated to be $\sim 3 \times 10^{4} -  10^{5}$~yr 
\citep{Andre00, Evans09}, compared to $\sim 2 - 5  \times  10^{5}$~yr for the Class~I phase \citep{Greene94, Evans09}. 
Moreover, regardless of their precise age and lifetime, Class~0 objects are envelope-dominated protostars 
($M_{env} >> M_\star $ -- \citealt{Andre93,Andre00}), while Class~I objects tend to have only residual protostellar envelopes 
($M_{env} < M_\star $ -- \citealt{Andre94a,Motte01a}). 
Therefore, Class~0 protostars are much more likely than Class~I objects to retain 
detailed information about the collapse initial conditions and the physics of the binary fragmentation process. 
Furthermore, the circumstellar mass reservoir left around Class~I sources is generally not sufficient to form companions more massive 
than substellar objects at the Class~I stage. 
Probing the multiplicity on scales $<$1000~AU as soon as the Class~0 stage is thus one 
of the keys to understanding the bulk of multiple star formation.

Only subarcsecond mm/submm interferometry can probe the inner structure of Class~0 objects, so that little is known 
about their multiplicity on scales $< 150$~AU. 
Several interferometric studies discovered a number of wide ($\simgt 1000 $~AU) multiple Class~0 systems  \citep{Looney00, Bourke01, Chen08, Girart09}, 
but these studies were limited by sensitivity to small samples of relatively luminous objects. 
The SMA has recently been used to study several Class~0 sources at 1.3~mm and 0.8~mm but only at  $\sim 2\arcsec $ ($\sim 450$~AU) resolution \citep{Jorgensen07b}, which does not allow the detection of tight multiple systems.
Among the eight Class~0 sources targeted with the SMA, only one source already embedded in a wide separate-envelope system (NGC1333-IRAS4A / NGC1333-IRAS4B) was found to exhibit a higher degree of multiplicity on smaller scales: \citet{Jorgensen07b} showed that IRAS~4A splits into two components separated by 
$\sim$2\arcsec ~(450~AU) and IRAS~4B splits into two components separated by $\sim$11\arcsec ~(2400~AU). 
These two new companions were detected at both 1.3~mm and 0.8~mm.

To determine the typical outcome of protostellar collapse and constrain binary fragmentation models, high-resolution 
imaging of very young protostars observed as early as possible after the end of the first collapse phase are crucially needed.
In this paper, we present the results of a pilot high-resolution survey of 5 Class 0 objects carried out with  
the IRAM Plateau de Bure Interferometer (PdBI) equipped with new-generation 1.3~mm receivers and 
using the most extended baselines of the interferometer (new A configuration).

\section{Observations and data reduction}

\subsection{Sample selection}

In order to probe the multiplicity at the Class~0 stage, we conducted a pilot survey of 5 sources with the Plateau de Bure Interferometer (PdBI) in February 2008.
The sources were first selected based on a distance criterion: they had to be close enough so that the 
PdBI resolution in the most extended configuration would probe $\simlt$150 AU scales, i.e. all sources had to be located at $d < 250-300$~pc. The second criterion was the source locations in the sky: the sources had to be visible from the PdBI in winter and observable with a synthetized beam  of less than 0.6$\arcsec$ (in both directions) with the A array. 
The selected sample includes the following five sources:
IRAM~04191, L1527, L1521F, all located in Taurus at $d \sim 140$~pc, and L1448-C and NGC~1333-IRAS2A both located in the Perseus cloud at $d \sim$ 250~pc (see Table~\ref{tab:sources}).
These 5 sources have M$_{env} \sim $~0.5--4~M$_{\odot}$, L$_{bol} \simlt$ 10 L$_{\odot} $ 
and are among the youngest known solar-type Class~0 protostars \citep{Andre00}. 
The main properties of the 5 Class~0 target sources are summarized below.

%%%%%%%%%%%
\begin {center}
\begin{table*}[!t]
\centering \par \caption{Class~0 source sample}
\begin{tabular}{lccccccc}
\hline
\hline
 {Source}  & $\alpha$ (J2000)& $\delta$ (J2000)& SFR & Distance & M$_{\rm{env}}$ & L$_{\rm{bol}}$ & Ref.\\
 {} & {} & {} & {} & (pc) & (M$_{\odot}$) & (L$_{\odot}$) & \\
 {} & {} & {} & {$^{(1)}$} & {$^{(2)}$} & & & {$^{(3)}$}\\
\hline
{L1448-C} & 03:25:38.87 & $+$30:44:05.4  & Perseus & 250 & 1.6 & 5 & (a) \\
{NGC~1333-IRS2A}& 03:28:55.58 & $+$31:14:37.1 & Perseus & 250 & 1.7 & 10 & (b)\\
{IRAM~04191}& 04:21:56.91 & $+$15:29:46.1 & Taurus & 140 & 0.5 -- 1.5 & 0.1 & (c) \\
{L1527}& 04:39:53.90 & $+$26:03:10.0 & Taurus & 140 & 0.8 -- 1.7 & 1.6& (a), (d) \\
{L1521-F} & 04:28:38.99 & $+$26:51:35.6 & Taurus & 140 & 0.7 -- 4 & 0.1 & (e), (f)\\
\hline
\hline
\end{tabular}
\begin{list}{}{}
\item[$^{(1)}$]{Star Forming Region with which the Class~0 object is associated}
\item[$^{(2)}$]{Recent estimates of the distance to Perseus range from 220 to 350 pc. 
Throughout this paper, we adopt a distance of 250 pc for the Perseus molecular cloud \citep{Enoch06}}.
\item[$^{(3)}$]{References for the adopted values of M$_{\rm{env}}$ and L$_{\rm{bol}}$: (a) \citet{Motte01a}, (b) \citet{Jorgensen07b}, (c) \citet{Andre99}, (d) \citet{Ohashi97a}, (e) \citet{Bourke06}, (f) \citet{Crapsi04}.}
\end{list}
\label{tab:sources}
\end{table*}
\end {center} 
%%%%%%%%%%% 
%

\subsubsection{L1448-C}
L1448-C is located in the Perseus molecular cloud, and was first detected as a 2 cm radio continuum source \citep{Curiel90}, associated with a strong millimeter continuum source \citep{Bachiller91c}. 
This is a well-known low-mass Class~0 protostar \citep{Barsony98} driving a powerful, highly 
collimated outflow (\citealt{Bachiller90}), which has been imaged at high resolution (synthetized HPBW beam $\sim$2.5\arcsec at $\sim$90~GHz) in CO and SiO 
with PdBI  \citep{Guilloteau92, Bachiller95c}. 
The molecular jet has been extensively studied since its discovery, and is known to exhibit very high-velocity features 
($\pm$70 km s$^{-1}$ -- \citealt{Bachiller90, Bachiller95c}).
\subsubsection{NGC~1333-IRAS2A}
NGC~1333-IRAS2A was first identified in 450~$\mu$m and 850~$\mu$m continuum observations \citep{Sandell94, Sandell01}, and is also detected at cm-wavelengths \citep{Rodriguez99, Reipurth02} and as a compact 3~mm continuum source \citep{Jorgensen04a}.
CO maps of the IRAS2 region show two outflows, directed north-south and east-west \citep{Liseau88, Knee00}, both originating near IRAS2A. 
Therefore, it has been argued that IRAS2A may be an unresolved protobinary.
However, neither the 2.7~mm BIMA observations of \citet{Looney00} nor the 1.3~mm SMA observations of \citet{Jorgensen07b} detect a companion to the source, despite a $\sim$3~mJy/beam sensitivity in both cases and beam sizes of 0.6\arcsec ~and 2.2\arcsec, respectively.
\subsubsection{IRAM~04191}
The very low luminosity Class 0 object, IRAM~04191$+$1522 (hereafter IRAM~04191), is located in the southern part of the Taurus molecular cloud and was originally discovered in the millimeter dust continuum \citep{Andre99}. Follow-up observations revealed the presence of a CO bipolar outflow
and a weak 3.6 cm VLA radio continuum source located at its center of symmetry  \citep{Andre99}, 
as well as extended infall and rotation motions in a prominent, flattened envelope \citep{Belloche02}. 
It is associated with a weak {\it{Spitzer}} source and has an estimated accretion luminosity of only 
$L_{int} \sim 0.08\, L_\odot $ \citep{Dunham06}. 
\subsubsection{L1527}
L1527 IRS (hereafter L1527), located in the Taurus molecular cloud, has been classified as a borderline Class 0/I object. 
It is observed in a nearly edge-on configuration ($\sim$90$^{\circ}$ viewing angle) \citep{Ohashi97a} and features  
a large, dense circumstellar envelope \citep{Ladd91, Chen95, Motte01a}.
It also exhibits a prominent bipolar outflow whose lobes are oriented along the east-west direction \citep{Parker91}. 

\subsubsection{L1521-F}
L1521-F, located in the Taurus molecular cloud, was originally classified as an evolved starless core 
\citep{Codella97, Onishi99a, Crapsi04}.
The high central density and infall asymmetry seen in the HCO$^{+}$(3--2) line indicate an object in the earliest stages of gravitational collapse \citep{Onishi99a}. $^{12}$CO (2--1) observations show no clear evidence of bipolar outflow emission. Recent {\it{Spitzer}} observations of L1521-F detected a low luminosity protostar at mid-infrared wavelengths ($>$ 5$\mu$m) and in the MIPS 160~$\mu$m data, which led to a reclassification of this object as a very low luminosity Class~0 object 
(e.g. \citealt{Bourke06, Terebey09}). 
Only scattered light is detected at near-infrared wavelengths with IRAC, in the form of a bipolar nebula oriented 
east-west which is probably tracing an outflow cavity.

\subsection{IRAM Plateau de Bure observations}

Observations of the five sources were carried out at 1.3~mm with the IRAM Plateau de Bure Interferometer (PdBI), equipped with new-generation receivers 
in February 2008 (PdBI project R068).
Broad band continuum emission and $^{12}$CO(2--1) emission were observed simultaneously, with the PdBI in its most extended configuration 
(A array with 6 antennas, providing 15 instantaneous baselines ranging from 24 m to 760 m). 
The proximity of the Taurus and Perseus clouds in the sky allowed us to use the same gain calibrators for the two regions, 
and therefore time-share two tracks of $\sim$10~hr on the five sources. Each track was divided unequally, depending on the expected fluxes of the sources. In particular, a factor of three more time was spent integrating on IRAM~04191 and L1521-F than on L1448-C.
Several nearby phase calibrators (mainly 0415$+$379 and 0528$+$134) were observed to determine the time-dependent complex antenna gains. 
The correlator bandpass was calibrated on the strong quasars 3C273 and 3C454.3, while the absolute flux density scale was derived from MWC349 and 3C84. The absolute flux calibration uncertainty is estimated to be $\sim$15$\%$. 
During the observations, one spectral unit of the correlator was tuned to the $^{12}$CO (2--1) line at 230.538 GHz. 
%
%%%%%%%%%%%
\begin {center}
\begin{table*}[!t]
\centering \par \caption{Rms noise levels and naturally-weighted beam sizes of the final maps}
\begin{tabular}{l|cc|cc|cc|cc}
\hline
\hline
 {Source}  &  \multicolumn{2}{|c|}{230GHz} & \multicolumn{2}{c|}{230GHz combined $^{(1)}$} & \multicolumn{2}{c}{$^{12}$CO(2--1)} &  \multicolumn{2}{|c}{107 GHz}  \\
 {} & {HPBW} & {rms} & {~~HPBW} & {~~rms} & {~~HPBW} & {rms~~} & {~~HPBW} & {rms~~} \\
 {} & {} & {(mJy/beam)} & {} & {(mJy/beam)} & {} & {(mJy/beam)} & {} & {(mJy/beam)} \\
\hline
{L1448-C} & 0.48$\arcsec \times$ 0.27$\arcsec$  &  0.9 & 1.68$\arcsec \times$ 1.39$\arcsec$ & 2.8  & 0.48$\arcsec \times$ 0.27$\arcsec$ & 10 & 4.08$\arcsec \times$ 3.27$\arcsec$ & 10 \\
{NGC~1333-IRS2A}& 0.49$\arcsec \times$ 0.32$\arcsec$ & 1.16 & - & - & 0.49$\arcsec \times$ 0.32$\arcsec$ & 20 & - & -  \\
{IRAM~04191}& 0.56$\arcsec \times$ 0.31$\arcsec$ & 0.37 & 1.37$\arcsec \times$ 0.82$\arcsec$ & 0.31 & 0.56$\arcsec \times$ 0.31$\arcsec$ & 10 & - & -  \\
{L1527}& 0.48$\arcsec \times$ 0.28$\arcsec$ & 1.2 & 0.88$\arcsec \times$ 0.79$\arcsec$ & 2.0 & 0.48$\arcsec \times$ 0.28$\arcsec$ & 20 & - & -  \\
{L1521-F} & 0.49$\arcsec \times$ 0.27$\arcsec$ & 0.12 & - & - & 0.48$\arcsec \times$ 0.27$\arcsec$ & 8 & - & -  \\
\hline
\hline
\end{tabular}
\begin{list}{}{}
\item[$^{(1)}$]{L1448-C and L1527: combination of 230~GHz data from project R068 with 218 GHz data from project G080. \\ 
~~IRAM~04191: combination of 230 GHz data from project R068 with 227 GHz data from Belloche et al. (2002).}
\end{list}
\label{tab:obs}
\end{table*}
\end {center} 
%%%%%%%%%%%
%
The total bandwidth of this $^{12}$CO spectral unit was 160 MHz, with individual channel spacings of 625 kHz (corresponding to a velocity resolution of 1.62 km s$^{-1}$). 
The remaining six windows of the correlator were combined to observe the continuum emission with a total bandwidth of 1.92 GHz between 229.5~GHz and 231.5~GHz. The average system temperature of the 1~mm receivers was $\sim$250 K. 
The typical angular resolution  
was 0.5$\arcsec \times$ 0.3$\arcsec$ (HPBW) at the declinations of the targets, 
while the full width at half maximum (FWHM) of the PdBI primary beam is $\sim$ 22$\arcsec$ at 230~GHz. 

\noindent In this study, we also make use of the 1.4~mm PdBI observations of both L1448-C and L1527, and the 2.8~mm observations of L1448-C, carried out in the B, C, D configurations between November 1996 and September 1998 (unpublished PdBI project G080 by Motte et al.).  
The typical resolution of these early PdBI observations was $\sim$4$\arcsec$ at 2.8~mm (107~GHz) and $\sim$2$\arcsec$ at 1.4~mm (219~GHz).
\\In addition, the 227~GHz observations obtained by \citet{Belloche02} toward IRAM~04191 in the B, C, D configurations of PdBI, 
which had an HPBW angular resolution $\sim$2$\arcsec$, were also used in combination with our A-configuration observations at 230~GHz.

\subsection{Interferometric data reduction}

%%%%%%%%
\begin{figure*}[!ht]
\begin{center}
\subfigure{\includegraphics [width=0.34\textwidth,angle=-90,trim=0cm 0cm 0cm 0cm,clip=true]{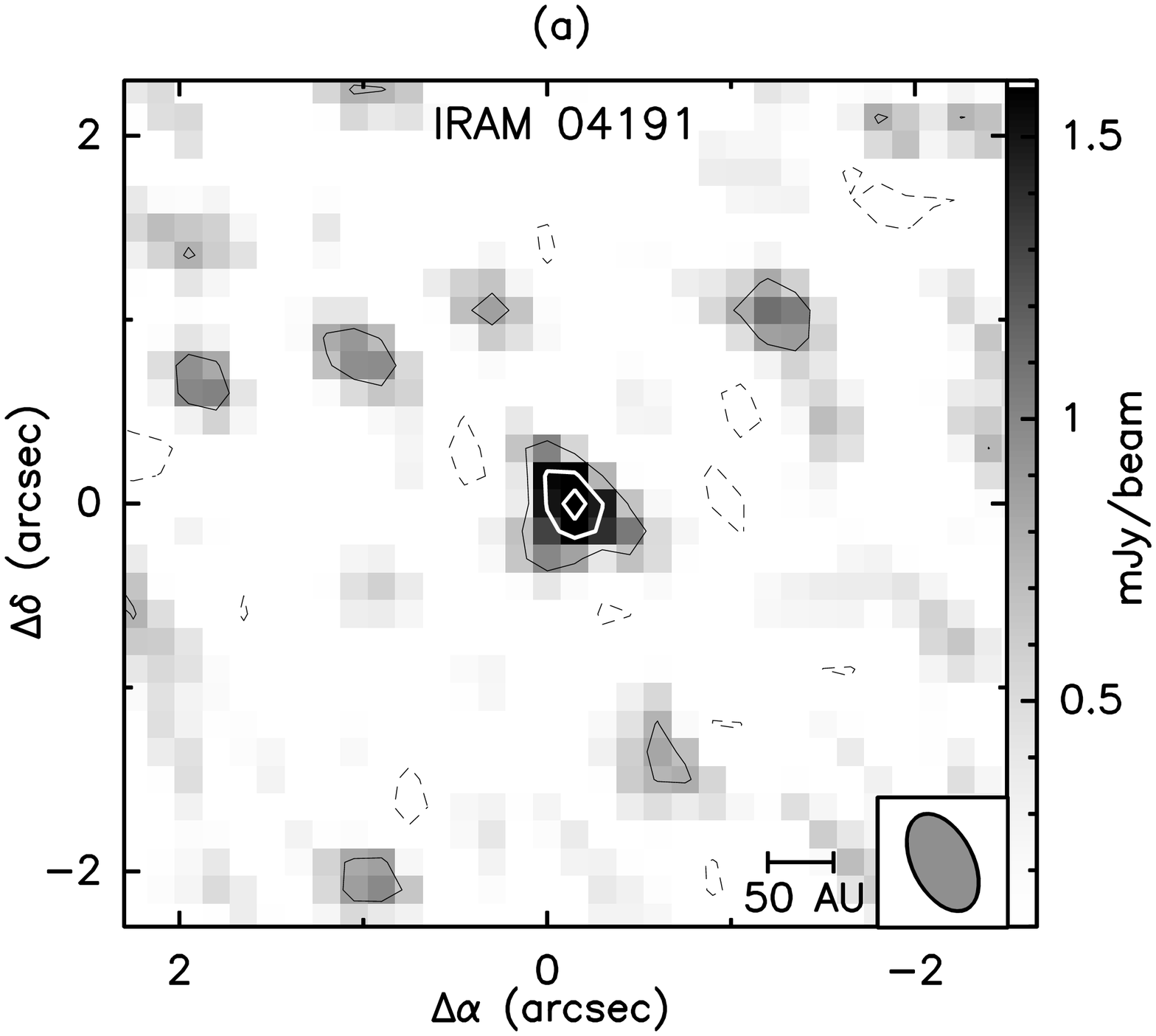}}
\hfill
\subfigure{\includegraphics [width=0.34\textwidth,angle=-90,trim=0cm 0cm 0cm 0cm,clip=true]{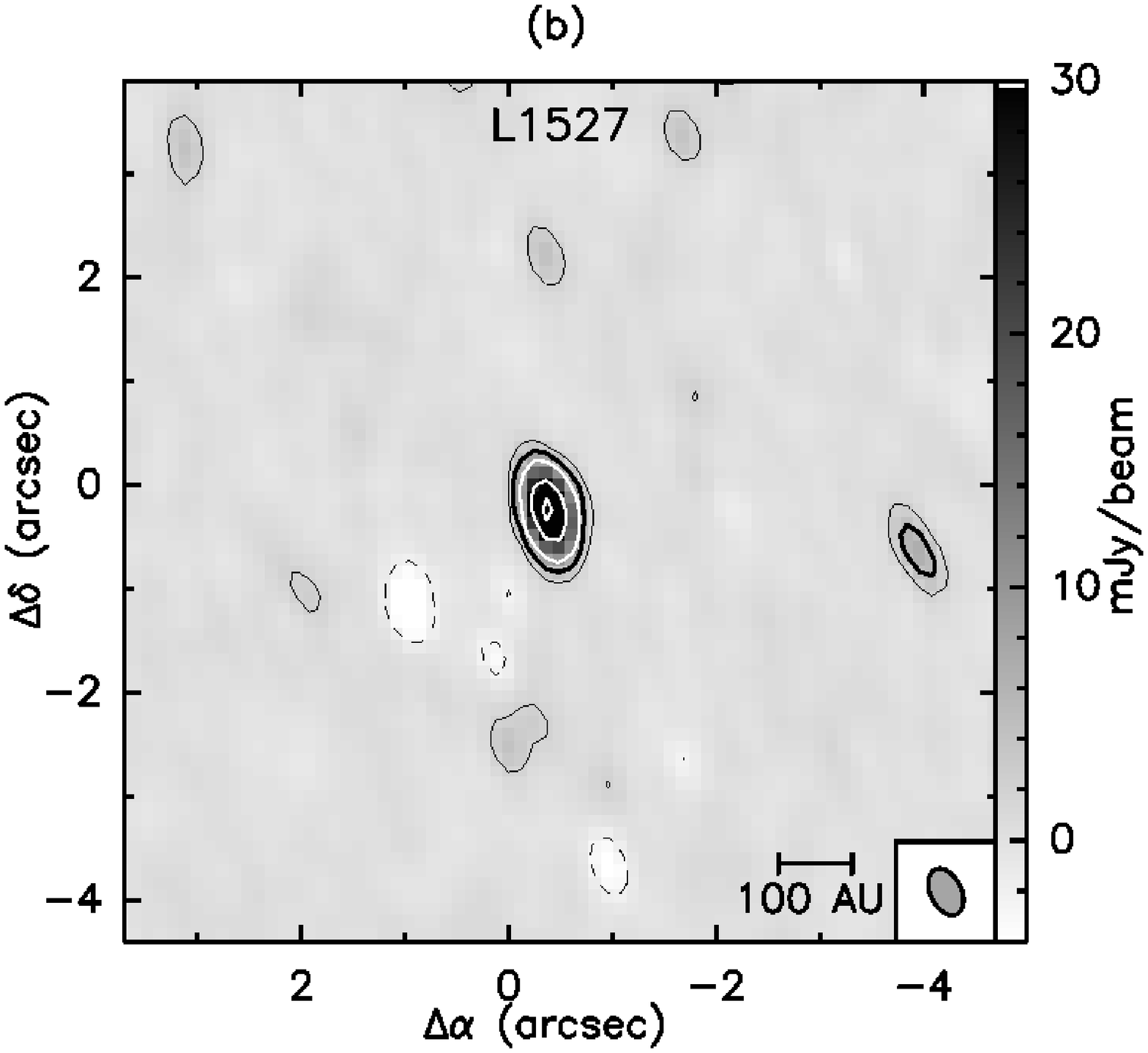}}
\\
\subfigure{\includegraphics [width=0.34\textwidth,angle=-90,trim=0cm 0cm 0cm 0cm,clip=true]{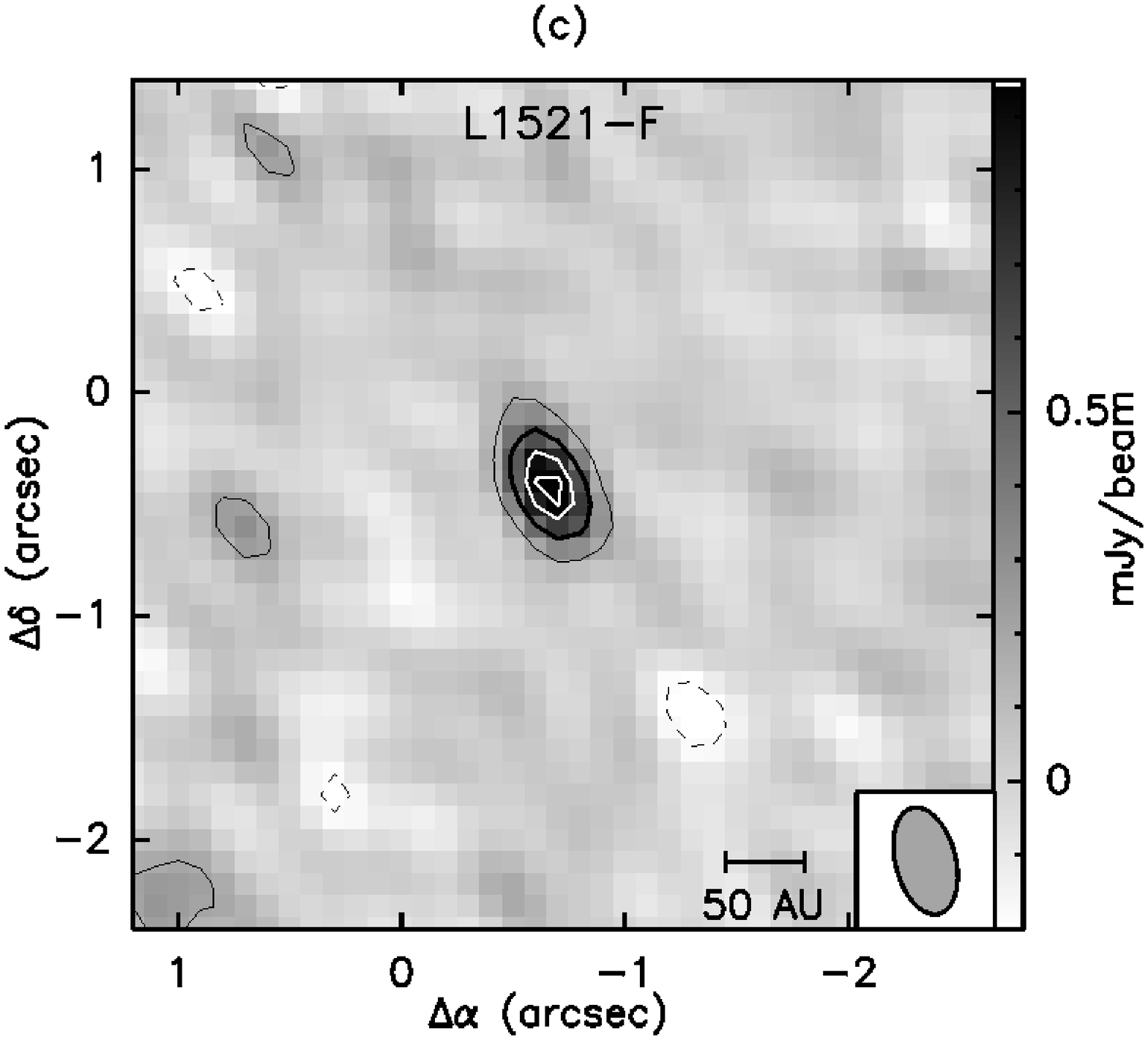}}
\caption{ High resolution 1.3~mm continuum maps of the Taurus sources. In all panels, the filled ellipse in the bottom right corner shows the synthetized HPBW beam.
{\it{(a)}} IRAM~04191. The synthetized HPBW is 0.57$\arcsec \times$ 0.33$\arcsec$, and the rms noise is $\sigma$ $\sim$ 0.37 mJy/beam. The contour levels are $-2\sigma$ (dashed), 2$\sigma$, 4$\sigma$ and 5$\sigma$.  
{\it{(b)}} L1527. The HPBW is 0.48$\arcsec \times$ 0.28$\arcsec$, and $\sigma$ $\sim$ 1.2 mJy/beam. The contour levels are $-2\sigma$ (dashed), 2$\sigma$ and 5$\sigma$ (bold), and 10$\sigma$, 30$\sigma$, 50$\sigma$ in white.  
{\it{(c)}} L1521-F. The HPBW is 0.49$\arcsec \times$ 0.27$\arcsec$, and $\sigma$ $\sim$ 0.1 mJy/beam. The contour levels are $-2\sigma$ (dashed), 2$\sigma$, 
5$\sigma$ (bold), and 8$\sigma$, 10$\sigma$ in white. 
}
\label{fig:R068_taurus}
\end{center}
%%%%%%%
%
%
%%%%%%%%
%\begin{figure*}[!h]
\begin{center}
\subfigure{\includegraphics [width=0.34\textwidth,angle=-90,trim=0cm 0cm 0cm 0cm,clip=true]{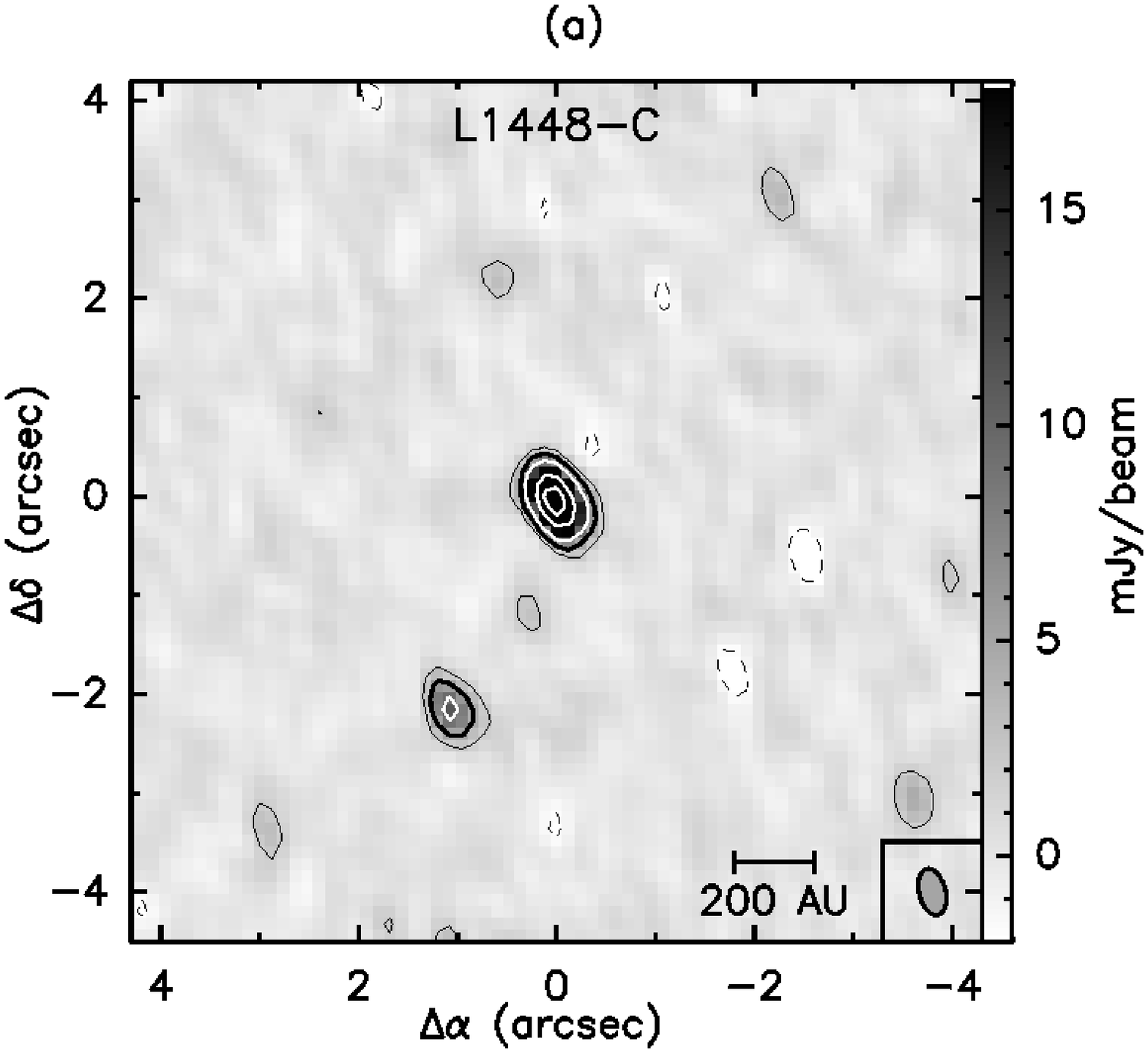}}
\hfill
\subfigure{\includegraphics [width=0.34\textwidth,angle=-90,trim=0cm 0cm 0cm 0cm,clip=true]{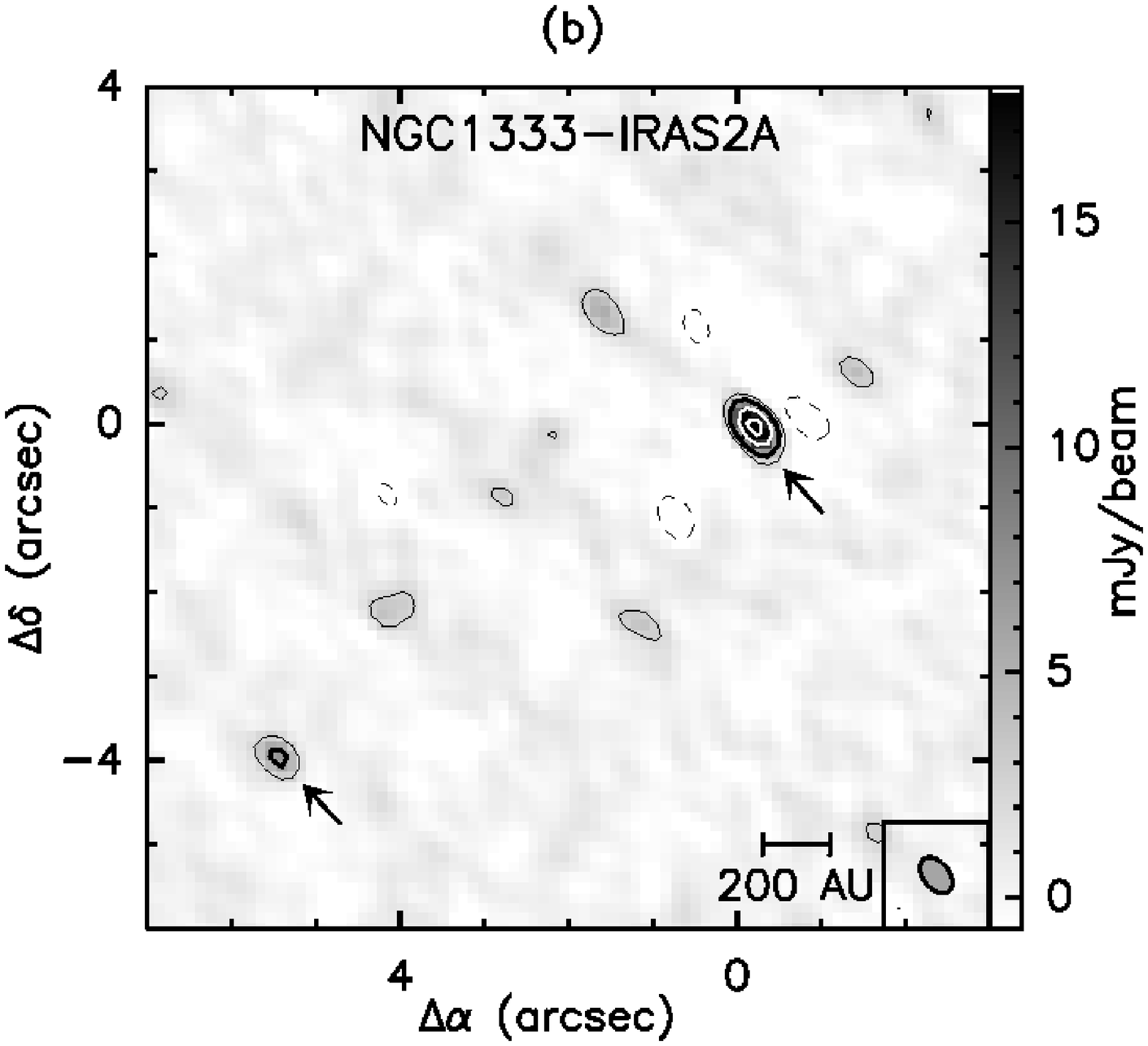}}
\caption{High resolution 1.3~mm continuum maps of the Perseus sources. In both panels, the filled ellipse in the bottom right corner indicates the synthesized HPBW beam. 
{\it{(a)}} L1448-C. The HPBW is 0.48$\arcsec \times$ 0.27$\arcsec$, and the rms noise is $\sigma$ $\sim$ 0.93 mJy/beam. The contour levels are -2$\sigma$ (dashed), 2$\sigma$, and 5$\sigma$ (bold), 10$\sigma$ to 50$\sigma$ by 20$\sigma$ in white.
{\it{(b)}} NGC~1333-IRAS2A. The HPBW  beam is 0.57$\arcsec \times$ 0.33$\arcsec$, and $\sigma$ $\sim$ 1.16 mJy/beam. The contour levels are -2$\sigma$(dashed), 2$\sigma$, 5$\sigma$ (bold) ; and 12$\sigma$, 20$\sigma$ in white. The two arrows show the positions of the two sources detected above the 5$\sigma$ level.}
\label{fig:R068_perseus}
\end{center}
\end{figure*}
%%%%%%%%

All the data were calibrated, mapped, and analyzed with the GILDAS\footnote{Grenoble Image and Line Data Analysis System, software provided and actively developed by IRAM (\textcolor{blue}{\url{http://www.iram.fr/IRAMFR/GILDAS}})} software package.
Each map was deconvolved down to the theoretical rms noise level using the MAPPING CLEAN method \citep{Clark80}. 
Natural weighting was applied to the measured visibilities, producing synthetized half power beam width (HPBW) resolutions $\sim$0.5$\arcsec \times$ 0.3$\arcsec$, as given in Table~\ref{tab:obs}. Note that the 3~mm continuum maps of L1448-C and L1527 have significantly 
larger synthetized beams (HPBW $\sim$ 4$\arcsec \times$ 3.5$\arcsec$) than the 1.3~mm maps, as no 3~mm data were obtained with the A-configuration. 
The restored continuum maps have rms values of 0.12 -- 2.8 mJy/beam (see Table~\ref{tab:obs}), depending on the integration time, array configuration and receivers used during the observations.
\\Likewise,  $^{12}$CO(2--1) data cubes were produced with natural uv-weighting, resulting in effective angular resolutions and rms noise values reported in the fourth column block of Table~\ref{tab:obs}. 
\\The 1.3~mm continuum visibilities obtained toward L1448-C and L1527 in the various PdBI configurations were merged together in order to produce high spatial dynamic range maps. 
Since the A-configuration data were obtained at a central frequency of 230.5 GHz, while the BCD-configuration data were obtained at a central frequency of 219 GHz, we had to scale the BCD-configuration data to 230.5 GHz assuming a spectral index $\alpha = 3$, corresponding to a dust emissivity index $\beta \sim \alpha -2 \sim 1$, 
for the emission of the inner protostellar environment (inner envelope $+$ disk) traced by our maps.
Therefore, the 219~GHz visibilities were scaled by $(230/219)^{3} \sim$1.16 before merging the R068 and G080 datasets at 230.5 GHz. 
The respective weights of each dataset were adjusted so as to produce the highest possible resolution image, while keeping a low rms noise. The synthetized beam sizes and rms sensitivities of the combined data are given in the third column block of Table~\ref{tab:obs}. 
The resulting maps are shown in Fig.~\ref{fig:l1448c_comb} and Fig.~\ref{fig:taurus_comb} below, respectively.
\\Following the same method, the 227~GHz PdBI visibilities taken by Belloche et al. (2002) toward IRAM~04191 
and the 230~GHz visibilities from our present A-configuration PdBI observations were also merged, in order 
to produce a high spatial dynamic range map of IRAM~04191 (see Table~\ref{tab:obs}) at 1.3~mm. 
The resulting map is shown in Fig.~\ref{fig:taurus_comb}a.

\section{Results of the A-configuration PdBI observations} 
\subsection{High-resolution 1.3~mm continuum maps}

%
%%%%%%%%
\begin {center}
\begin{table*}[!t]
\centering \par \caption{Properties of the 1.3~mm continuum sources detected in the high-resolution PdBI maps}
\begin{tabular}{l|cc|c|c|c|c}
\hline
\hline
 {Source}  &  \multicolumn{2}{|c|}{Position (1.3~mm)} & {Peak flux} & {FWHM $^{(1)}$} & {Flux density $^{(2)}$} & {Separation $^{(3)}$} \\
 {} & {$\alpha$ (J2000)} & {$\delta$ (J2000)} & {(mJy/beam)} & {(arcsec)} & {(mJy)} & {(AU)}\\
\hline
{L1448-C} & {03:25:38.87} & {30:44:05.3} & 59 $\pm$ 0.9& 0.9 $\pm$ 0.05  & 57 $\pm$ 2 & - \\
{L1448-C / South1}& {03:25:38.95} & {30:44:03.2} & 11 $\pm$ 0.7 & 0.7 $\pm$ 0.1 & 11 $\pm$ 2 & 600 \\
{NGC~1333-IRS2A}& {03:28:55.56} & {31:14:37.1} & 26 $\pm$ 1.16 & 0.55 $\pm$ 0.07 & 18 $\pm$ 2 & - \\
{NGC~1333-IRS2A / SE}&03:28:56.00 & 31:14:33.1 & 7 $\pm$ 1.16 & 0.23 $\pm$ 0.04 & 4 $\pm$ 1.5 & 1900 \\
{IRAM~04191}& 04:21:56.90 & 15:29:46.1 & 2 $\pm$ 0.37 & 0.28  $\pm$ 0.02 & 0.9 $\pm$ 0.4 & - \\
{L1527}& 04:39:53.87 & 26:03:09.8 & 50 $\pm$ 1.2 & 0.42 $\pm$ 0.05 & 65 $\pm$ 3 & - \\
{L1521-F} & 04:28:38.94 & 26:51:35.2 & 1 $\pm$ 0.12 & 0.65 $\pm$ 0.009 & 0.8 $\pm$ 0.2 & - \\
\hline
\hline
\end{tabular}
\begin{list}{}{}
\item[$^{(1)}$]{FWHM diameter is computed from a circular Gaussian fit to the visibilities}
\item[$^{(2)}$]{Flux integrated above the 3$\sigma$ level.}
\item[$^{(3)}$]{Projected distance to the primary source.}
\end{list}
\label{tab:r068_carac}
\end{table*}
\end {center} 
%%%%%%%%
%
%

\noindent The 1.3~mm dust continuum maps we obtained with the A-array of PdBI toward the 5 sources  are 
shown in Figures 1 and 2.
The effective spatial resolution of these maps is better than 70~AU and 125~AU  (HPBW)  for the Taurus and Perseus sources, respectively.
All five primary Class~0 targets are detected in these high-resolution maps, 
with signal-to-noise ratios ranging from 5.4 (IRAM~04191) to 65 (L1448-C) (see Table~\ref{tab:r068_carac} for flux densities and source sizes).

\noindent The three Class~0 sources of Taurus targeted in the present study are found to be single in our maps: the main protostellar object is the only source detected above the 5$\sigma$ level in each of the maps shown in  Fig.~1. 
A tentative 5$\sigma$ ($\sim 6.6 $~mJy/beam) secondary component is detected in the L1527 map  (3.6$\arcsec$ west of the main source), but it 
is located at the edge of a dirty lobe, which casts doubt on the detection. Moreover, this tentative component
is not detected in the combined map shown in Figure~\ref{fig:taurus_comb} (while such a source should have been detected, given the rms noise level of the combined map). Therefore, we will not mention this source anymore in the following. 
\\On the other hand, both Perseus maps (Figs.~2a and 2b) show evidence of secondary 1.3~mm continuum components  detected above the 5$\sigma$ level.
The map of L1448-C reveals a secondary 1.3~mm continuum source, located $\sim$2.4$\arcsec$ south-east from 
the primary L1448-C source (Table~\ref{tab:r068_carac}). It is the only additional component detected above the 3$\sigma$ level in the whole 22$\arcsec$ map (see Fig.~\ref{fig:R068_perseus} and Fig.~\ref{fig:cont_co}). 
The map of NGC~1333-IRAS2A (see Fig.~\ref{fig:R068_perseus}) also shows a secondary 1.3~mm continuum source 
(see position in Table~\ref{tab:r068_carac}) detected above the 5$\sigma$ level. 
The nature of these secondary sources is discussed further in Sect.~4 below.\\

\subsection{$^{12}$CO(2--1) data}
Compact $^{12}$CO(2--1) emission is detected toward all of the targets, except L1521-F. Moreover, significant high-velocity $^{12}$CO(2--1) emission is detected only toward L1448-C.
This is due to the fact that the A configuration of PdBI filters out most of the extended $^{12}$CO(2--1) emission from both 
protostellar outflows and the parent molecular clouds.

%%%%%%%%
\begin{figure*}[!ht]
\begin{center}
 \includegraphics[width=0.90\textwidth,angle=0,trim=0cm 0cm 0cm 0cm,clip=true]{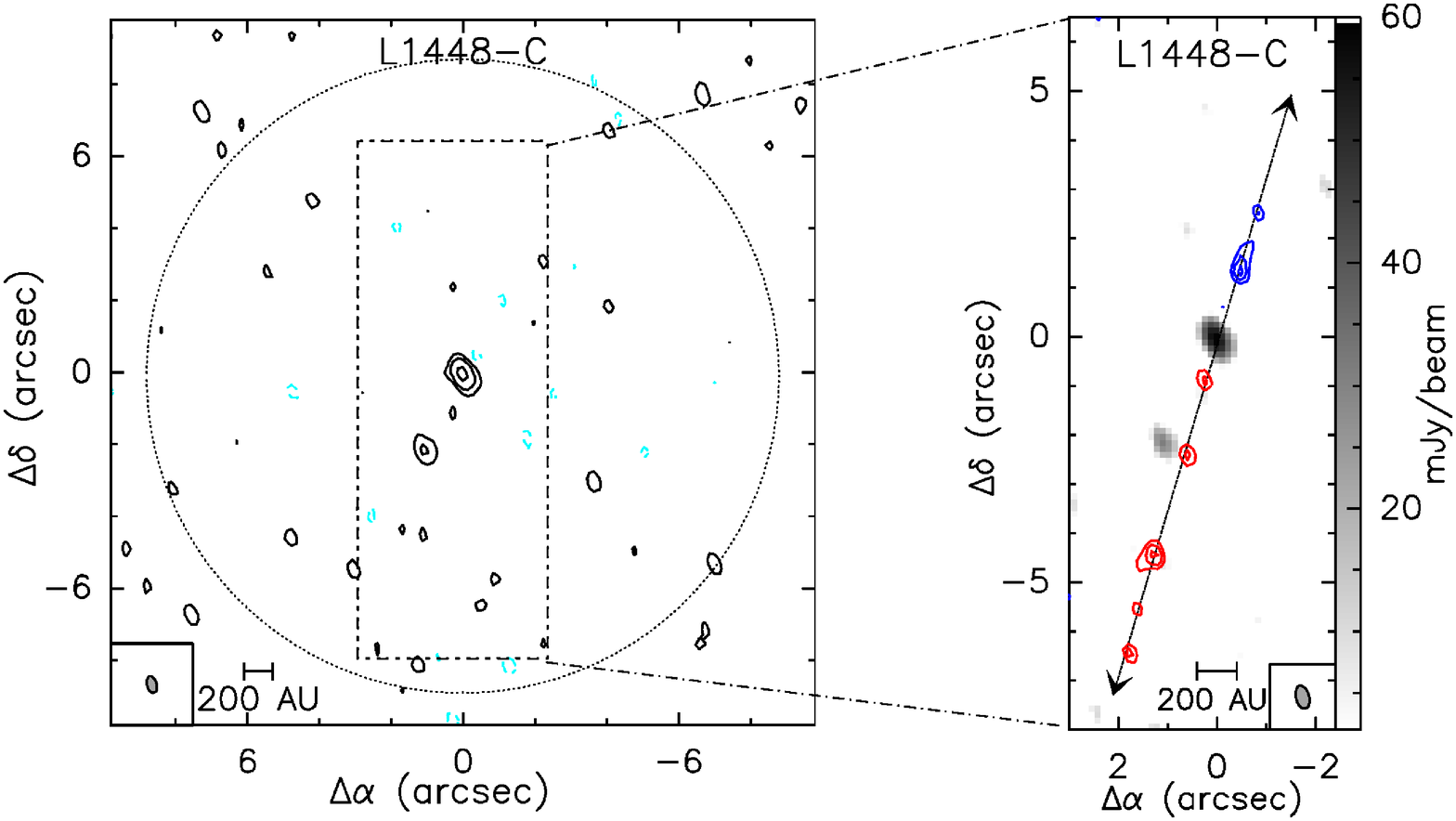}
\caption{The left panel shows the high-resolution 1.3~mm map of L1448-C (same as Fig.~\ref{fig:R068_perseus} but showing a wider area covering most of the primary beam). The large dotted circle represents the cleaned area of the map.
Note that the large-scale map has not been corrected for primary beam attenuation at large distances from the centre (${\it FWHM} = 22\arcsec$). The contour levels are $-2\sigma$ (light blue dashed), 2$\sigma$, 8$\sigma$ and 40$\sigma$. The right panel shows a blow-up of the central part of the high-resolution 1.3~mm map of L1448-C in greyscale, with superimposed contours of the high-velocity $^{12}$CO(2--1) emission integrated from $-$60 to $-$40~km s$^{-1}$ (blue contours) and from $+$50 to $+$80~km s$^{-1}$ (red contours). 
The first $^{12}$CO(2--1) contour corresponds to the 3$\sigma$ level (30~mJy/beam); the next contour levels are 5$\sigma$ and 10$\sigma$. The double arrow marks the direction of the high-velocity jet detected in $^{12}$CO(2--1). The filled ellipse at the bottom of the panels indicates the synthesized HPBW resolution.}
\label{fig:cont_co}
\end{center}
\end{figure*}
%%%%%%%%

\noindent In the $^{12}$CO(2--1) map of L1448-C, a total of seven compact high-velocity CO "bullets" are detected along the bipolar jet axis, in both the redshifted and blueshifted lobes (see Fig.~\ref{fig:cont_co}). 
Two blueshifted bullets are detected above the 3$\sigma$ level (30 K.km/s) at LSR velocities ranging from $-$60~km s$^{-1}$ to $-$40~km s$^{-1}$, and are located 1.44$\arcsec$ (360~AU) and 2.69$\arcsec$ (670~AU) away from the primary source driving the jet, respectively.
Five redshifted bullets are detected along the redshifted jet axis, south-west of the driving source.
These five features have LSR velocities ranging from 50~km s$^{-1}$ to 80~km s$^{-1}$, and are located at distances from the driving source ranging from 0.9$\arcsec$ (125~AU) to 6.7$\arcsec$ (940~AU).
The seven high-velocity features detected in the L1448-C map are all remarkably well aligned with the axis of the molecular jet already mapped at lower resolution by Bachiller et al. (1995) with PdBI, and more recently by Jorgensen et al. (2007) with SMA.
Therefore, we conclude that these high-velocity bullets trace the inner part of the jet driven by the protostellar source L1448-C.

\section{Nature of the secondary components detected in the millimeter continuum maps}

\subsection{Sources detected in the vicinity of L1448-C}

\subsubsection{Secondary 1.3~mm continuum source}

In the high resolution 1.3~mm continuum map of L1448-C shown in Fig.~\ref{fig:R068_perseus}a, a secondary source is detected $\sim$2.4$\arcsec$ south-east of the main source. In the combined 1.3~mm continuum map, this secondary component is no longer resolved from the primary source (see Fig.~\ref{fig:l1448c_comb}), because its peak flux (11.3 mJy/beam) is only between the 3$\sigma$ and 5$\sigma$ levels, and it is confused with extended 1.3~mm emission south-east of L1448-C, 
which likely arises from a cavity in the red-shifted outflow lobe.
This secondary source is not detected either in the 3~mm map shown in Fig.~\ref{fig:l1448c_comb}b due to insufficient angular resolution. 
But it lies close to the L1448-C jet axis  and is immediately adjacent  to the second redshifted high-velocity bullet detected in our $^{12}$CO(2--1) observations 
(only $\sim$0.5$\arcsec$ separation -- see Fig.~\ref{fig:cont_co}).
Furthermore, this secondary 1.3~mm continuum source coincides with the position of the first SiO(2--1) peak (clump RI) detected by \citet{Guilloteau92} in the red-shifted outflow lobe. This traces the presence of an outflow-induced shock at this position, with a high LSR velocity offset of +50~km s$^{-1}$ (see Fig.~\ref{fig:3mm_vs_sio}). 
Therefore, we conclude that the secondary 1.3~mm source detected south-east of L1448-C is not a genuine protostellar companion but rather an outflow feature directly associated with the high-velocity jet from the primary Class~0 object.

\subsubsection{Secondary 3~mm source and adjacent {\it{Spitzer}} source}

\begin{figure}[!h]
\begin{center}
\subfigure{\includegraphics[width=0.70\columnwidth,angle=0,trim=0cm 0cm 22.5cm 0cm,clip=true]{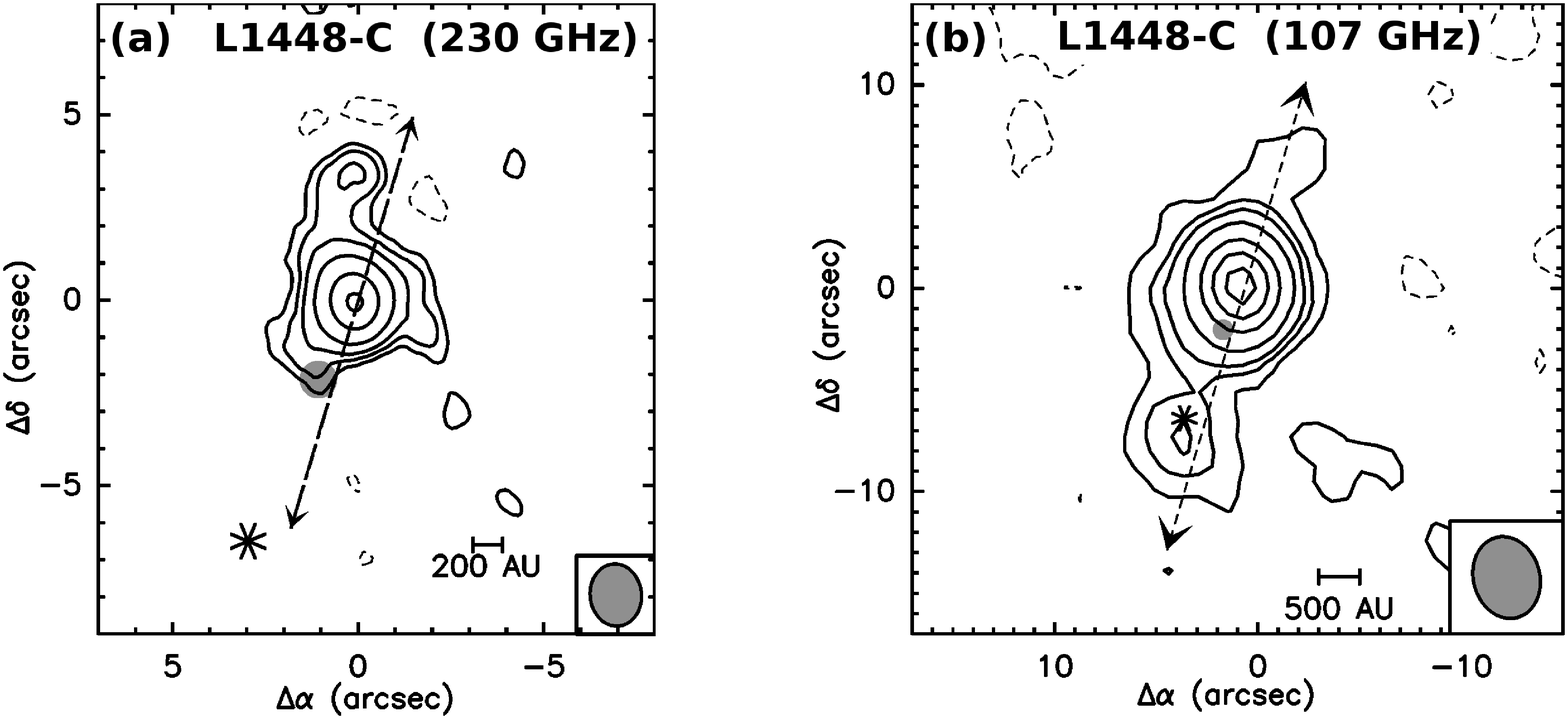}}\\
\subfigure{\includegraphics[width=0.75\columnwidth,angle=0,trim=20.5cm 0cm 0cm 0cm,clip=true]{13492_fig7.eps}}
\caption{
{\it{(a)}} Combined 1.3~mm dust continuum map of L1448-C. This image was constructed by combining the 230~GHz visibilities obtained with A-configuration of PdBI with the 219~GHz visibilities obtained with the B, C and D-configurations of PdBI. The synthetized HPBW is 1.68$\arcsec \times$ 1.39$\arcsec$, and the rms noise is $\sigma$ is $\sim$ 2.8 mJy/beam. The dashed contours are levels of -3$\sigma$. The full contours are levels of 3$\sigma$, 5$\sigma$, 8$\sigma$ and 15$\sigma$ to 45$\sigma$ by 15$\sigma$. 
The grey circle indicates the position of the secondary 1.3~mm continuum source detected in the high resolution map shown in Fig.~\ref{fig:R068_perseus}. The arrows indicate the direction of the high-velocity jet observed in $^{12}$CO(2--1) (see \S3.2).
{\it{(b)}} 107~GHz continuum map of L1448-C. The synthetized HPBW is 4.08$\arcsec \times$ 3.27$\arcsec$, and the rms noise is $\sigma \sim$0.49 mJy/beam.  
The dashed contour level is -2$\sigma$. The full contours are levels of 2$\sigma$, 5$\sigma$, 8$\sigma$ and 15$\sigma$ to 85$\sigma$ by 15$\sigma$. In both panels, the filled ellipses in the lower right corner indicate the synthesized HPBW. The star indicates the position of the {\it{Spitzer}} source reported by \citet{Jorgensen06}.}
\label{fig:l1448c_comb}
\end{center}
\end{figure}

Figure~\ref{fig:l1448c_comb}b shows a 3~mm (107~GHz) continuum map 
of L1448-C based on BCD-array data taken with PdBI in 1997 (see Sect.~2.2 for details). 
It reveals the presence of a secondary 3~mm continuum source, clearly detected above the 5$\sigma$ level at position (03h25m39.10s, +30$^{\circ}$43$\arcmin$57.8$\arcsec$), i.e. $\sim$8.4$\arcsec$ south-east of the main L1448-C source, and which does not have any significant 
1.3~mm counterpart in Figure~\ref{fig:l1448c_comb}a. 
A mid-infrared source was recently detected with Spitzer 0.6$\pm$ 0.2$\arcsec$ north of this secondary 3~mm source \citep{Jorgensen06}, 
and $\sim$7.8$\arcsec$ south of L1448-C.
The small angular separation between the PdBI 3~mm source and the {\it{Spitzer}} mid-infrared source suggests that they are physically related.
Neither of our 1.3~mm continuum maps shows counterparts to the {\it{Spitzer}} or the 3~mm source above the 3$\sigma$ level, 
corresponding to upper limits to the 1.3~mm peak flux density of 3~mJy/0.48$\arcsec \times$ 0.27$\arcsec$ beam in the high resolution 1.3~mm map of L1448-C, 
and 8.4~mJy/1.68$\arcsec \times$ 1.39$\arcsec$ beam in the combined 1.3~mm map. 
The non detection of  the 3~mm source at 1.3~mm implies that the spectral index of the emission is $\alpha^{\rm{1.3mm}}_{\rm{3mm}} < 2.0$ 
(after scaling the 1.3~mm and 3~mm fluxes to matching beams), excluding the possibility that the whole millimeter emission is due to dust continuum emission from an embedded protostellar object (in this case $\alpha = \beta + 2$, where $\beta$ is the dust emissivity index, and $\alpha^{\rm{1.3mm}}_{\rm{3mm}} \sim$ 3 -- 4 
is expected -- e.g. \citealt{Dent98}).
%
%%%%%%%%
%%%%%%%%
%
\\ If the 3~mm source corresponds to a protostellar object, its non-detection in the 1.3~mm continuum emission is unlikely to result 
from interferometric filtering since Class~0 sources have strongly centrally condensed envelopes \citep{Andre00,Motte01a}, 
and are thus expected to be detected in a $\sim$1.5$\arcsec$ beam.
For instance, the 1.3~mm peak flux density of L1448-C is $\sim$134 mJy/beam in a $\sim$1.5$\arcsec$ beam, so the 3~mm secondary source would have to be 17 times weaker at 1.3~mm on $\sim$375~AU scales to be undetected (above 3$\sigma$ level i.e. 8.4 mJy/beam), which is more than twice the flux ration $\sim$8 measured between the two sources computed at 3~mm.
To further test whether the {\it{Spitzer}} and 3~mm continuum emission may originate from a protostellar object, 
we compared the observations with the publicly available grid of model YSO spectral energy distributions (SEDs) 
published by \citet{Robitaille07}. 
We tried to reproduce both the mid-infrared fluxes derived by \citet{Jorgensen06} for the southern {\it{Spitzer}} source and the PdBI 3~mm flux, while keeping the 
1.3~mm flux density lower than the 5$\sigma$ detection level ($\sim$5~mJy in a 0.37$\arcsec$ radius aperture) achieved with PdBI.
None of the models explored in this way can reproduce the data points properly : the ten best models have total $\chi ^{2}$ values $\sim$ 100 -- 200 for five data points (compared to total $\chi^{2} \sim$ 20-40 with data at the same five wavelengths for L1448-C), and show 1.3~mm fluxes which should have been detected above the 3$\sigma$ level in our PdBI observations.
The mid-infrared emission detected with {\it{Spitzer}} and the adjacent 3~mm emission detected with PdBI 
are therefore unlikely to originate from a protostellar object.
%
%%%%%%%%
\begin{figure}[!h]
\begin{center}
  \subfigure{\includegraphics[width=0.70\columnwidth,angle=0,trim=0cm 0cm 0cm 0cm,clip=true]{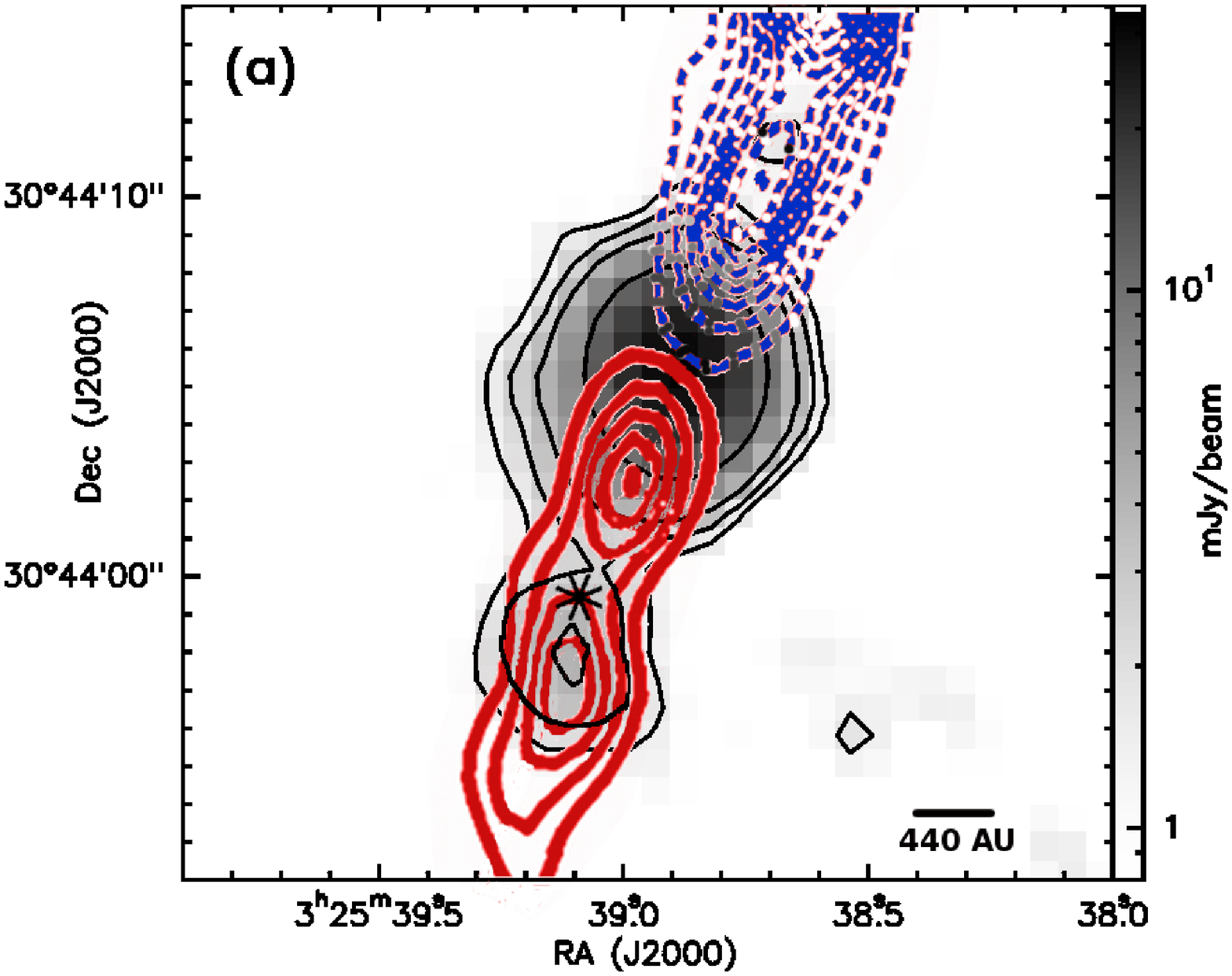}}
  \\
   \subfigure{\includegraphics[width=0.67\columnwidth,angle=0,trim=0cm 0cm 0cm 0cm,clip=true]{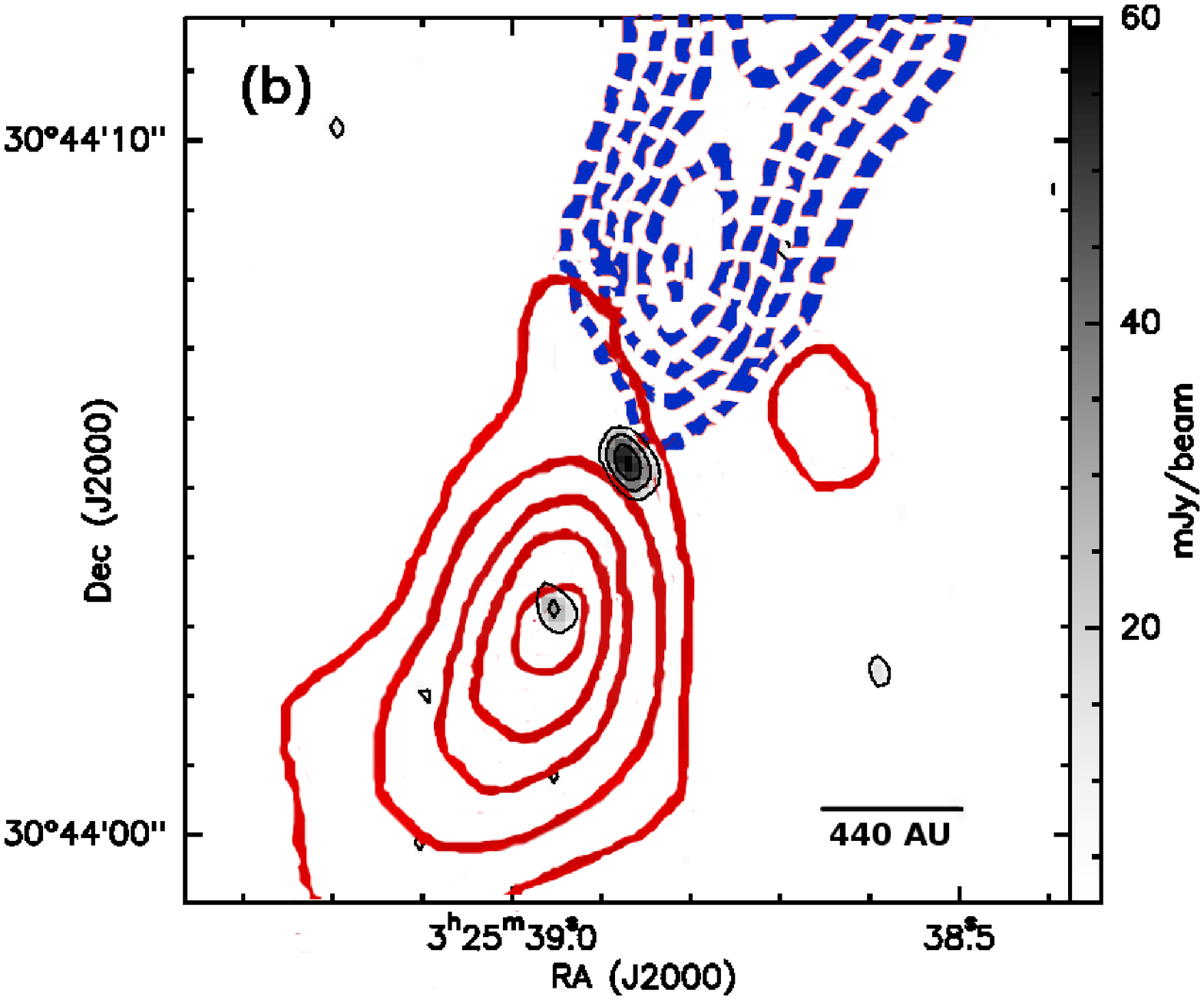}}
\caption{{\it{(a)}} Image and black contours show the 3~mm continuum map of L1448-C (same as Fig.~\ref{fig:l1448c_comb}). The red and blue contours are levels of the SiO(2--1) line intensity at $\pm$65 km s$^{-1}$ from \citet{Guilloteau92}. The star indicates the position of the {\it{Spitzer}} source reported by Jorgensen et al. (2006), for which no counterpart is detected above the 3$\sigma$ level in the 1.3~mm map (see \S4.1 for a discussion).
{\it{(b)}} Image and black contours show the high resolution 1.3~mm map of L1448-C. The red and blue contours are levels of SiO(2--1) line intensity at $\pm$50 km s$^{-1}$. 
One can see that both secondary (1.3~mm and  3~mm) sources coincide with peaks of SiO(2--1) emission.}
\label{fig:3mm_vs_sio}
\end{center}
\end{figure}
%%%%%%%%
%
\\The 3~mm and {\it{Spitzer}} sources are both located in the walls of an outflow cavity (see Jorgensen et al, 2007). They both coincide with the second SiO(2--1) emission peak detected toward the red-shifted lobe (clump RII in \citealt{Guilloteau92}) which reveals the presence of an outflow-induced shock at this location, at a high LSR velocity offset of +65~km s$^{-1}$ (see Fig.~\ref{fig:3mm_vs_sio}).
Furthermore, these two sources also coincide with a peak in the NH$_{3}$(2,2)/NH$_{3}$(1,1) ratio, which traces heating due to the interaction between the energetic outflow and the ambient molecular gas \citep{Curiel99}. 
We therefore propose that these two adjacent sources, located along the jet axis, are in fact shock-generated 
features \citep[cf.][]{Hartigan03} associated with the powerful outflow driven by L1448-C. 
Compact mid-infrared continuum emission along protostellar jets has already been observed: \citet{Lefloch05} reported the detection of such features along the HH~2 protostellar jet, and argued that the mid-IR emission arises from heating of very small grains formed by evaporation of dust grain mantles in outflow-induced shocks. 
Furthermore, the lack of a 1.3~mm counterpart to the 3~mm source can be explained by the nature of the 3~mm emission. 
The non-detection of VLA 2~cm emission by \citet{Curiel90} at the position of the 3~mm emission implies a spectral index $\alpha^{\rm{3mm}}_{\rm{2cm}} > 1.5$. 
Combined with $\alpha^{\rm{1.3mm}}_{\rm{3mm}} < 2.0$, this suggests that the 3~mm continuum emission is a combination 
of optically thick free-free emission {($1 \simlt \alpha^{\rm{3mm}}_{\rm{2cm}} \leq 2$ -- e.g. \citealt{Ghavamian98}) 
and optically thin dust continuum emission ($\alpha^{\rm{1.3mm}}_{\rm{3mm}} \sim $~ 2 -- 4) associated with a shock in the L1448-C outflow.
The fact that this outflow-induced shock (traced by the {\it{Spitzer}} and 3~mm sources) is not detected at 2~$\mu$m can be ascribed to high visual extinction toward the southern lobe ($A_{\rm{V}} \sim 32$ -- \citealt{Dionatos09}).
\\ Based on this multiwavelength analysis, we conclude that the {\it{Spitzer}} mid-infrared source and the PdBI 3~mm source detected $\sim$8$\arcsec$ south of L1448-C do not correspond to protostellar objects but are both tracing heating and compression resulting from an oblique shock on the outflow cavity wall.

\subsection{Secondary 1.3~mm continuum source detected near NGC~1333-IRAS2A}
Our high resolution PdBI observations of NGC~1333-IRAS2A allow us to probe the circumstellar environment of this source down to $\sim$90~AU scales, with better sensitivity than that achieved in previous interferometric observations. 
While no secondary component is detected above the 2$\sigma$ level within a radius  of 1000~AU ($\sim$4\arcsec) from the primary source, 
a new secondary 1.3~mm component is detected 7.7$\arcsec$ ($\sim$1900~AU) south-east of NGC~1333-IRS2A, 
with a peak flux of $\sim$7 mJy/beam (see  Fig.~\ref{fig:R068_perseus}b). 
This secondary 1.3~mm source is located near the east-west outflow originating from the vicinity of IRAS2A (see Fig.~\ref{fig:ngc1333_out}), 
but is not associated with any known high-velocity CO bullet or shock feature. Therefore, the nature of this 
source is unclear: it may either be a genuine protostellar companion or an outflow feature.
We name this secondary source NGC~1333-IRS2A/SE, as it probably lies within the same envelope as NGC~1333-IRS2A. 
The weakness of this source explains its non-detection in previous millimeter interferometric studies \citep{Looney00, Jorgensen07b}.

%%%%%%%%
\begin{figure}[!h]
\begin{center}
\includegraphics[width=0.7\columnwidth,angle=0,trim=0cm 0cm 0cm 0cm,clip=true]{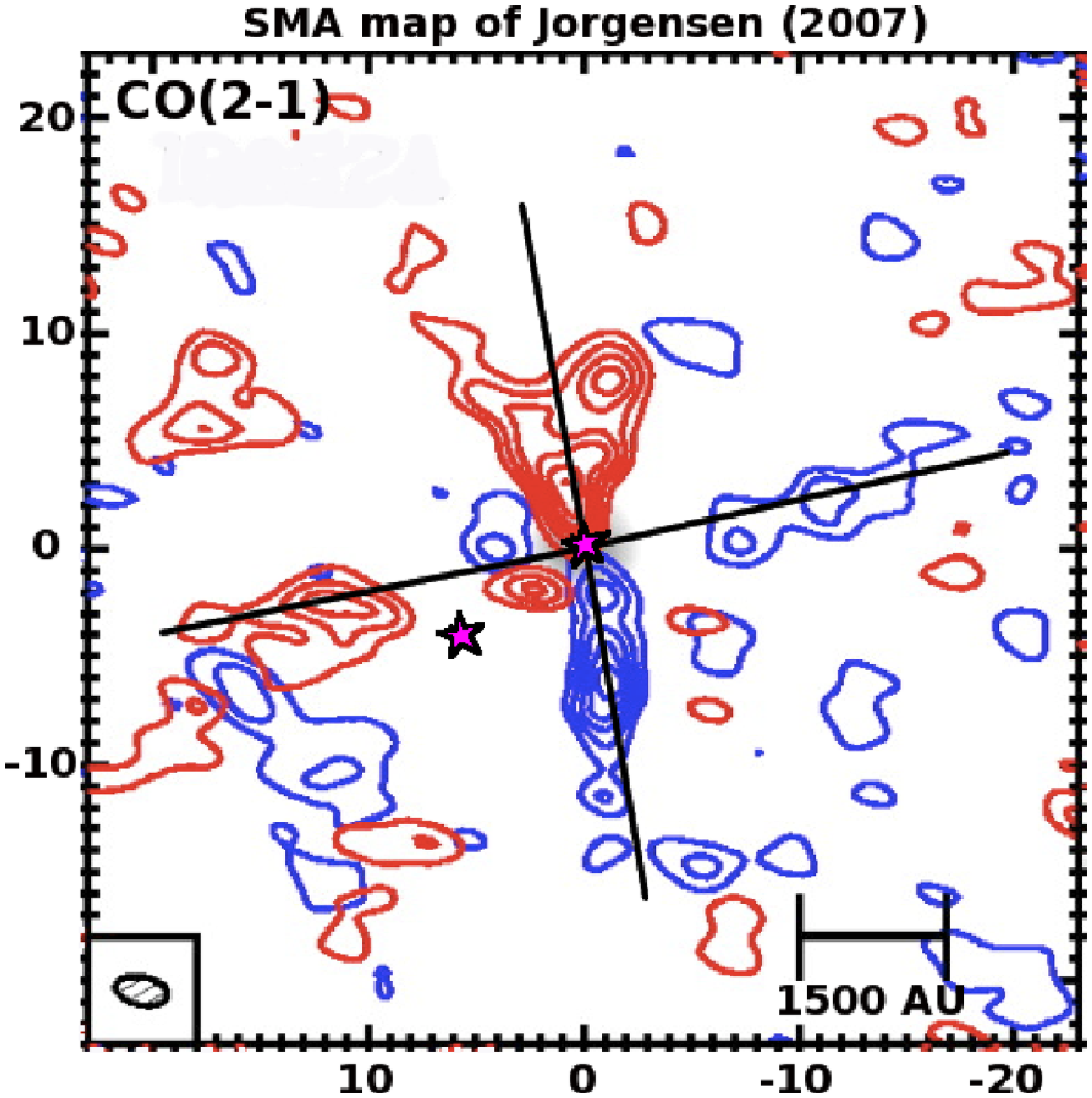}
\caption{SMA map  obtained in $^{12}$CO(2--1) by \citet{Jorgensen07b} toward NGC~1333-IRS2A. 
The color contours show $^{12}$CO(2--1) integrated intensity levels in steps of 3$\sigma$ (see \citet{Jorgensen07b} for details), with the blue contours indicating emission integrated from -6 to -1 km s$^{-1}$ relative to the systemic velocity, and red contours emission integrated from +1 to +6 km s$^{-1}$ relative to the systemic velocity. The two lines mark the direction of the protostellar outflows originating near IRAS2A. On large scales, the $\sim$ east-west outflow is more collimated than the $\sim$north-south bipolar one, and was found to be less energetic than the north-south bipolar outflow \citep{Knee00}. The two stars mark the position of the 1.3~mm continuum sources detected in our PdBI observations.}
\label{fig:ngc1333_out}
\end{center}
\end{figure}
%%%%%%%%

\section{Dust continuum emission detected toward the primary Class~0 sources}
A detailed discussion of the small scale properties and detailed morphology of the dust emission detected toward the primary Class~0 sources is beyond the scope 
of this paper and will be the subject of a forthcoming paper. Here, we provide a simple, qualitative description of the slightly extended 1.3~mm continuum emission detected with PdBI toward the primary Class~0 sources of our sample. 
\\Our high-resolution 1.3~mm continuum maps exhibit only little extended emission. 
This can be explained by the very high resolution achieved in these interferometer maps and the lack of short-spacing data, which filters out most of the extended emission from the envelope material.
In the combined 1.3~mm and 3~mm maps, however, the use of multiple array configurations allows us to recover some of the extended emission on scales ranging from $\sim$100 AU to $\sim$500 AU. 
Note that, as the older PdBI observations toward L1448-C and L1527 (G080 data taken in the B, C, D configurations) are noisier, the combined 1.3~mm maps have an improved spatial dynamic range (facilitating, e.g., image reconstruction), but higher rms noise values. 
The combined 1.3~mm maps of the two sources IRAM~04191 and L1527 are shown in Fig.~\ref{fig:taurus_comb}a and Fig.~\ref{fig:taurus_comb}b, respectively,  
while the combined 1.3~mm map of L1448-C was already shown in Fig.~\ref{fig:l1448c_comb}a above. 
Details about the combination procedure can be found in Sect.~2.3 above. 

%%%%%%%%
\begin{figure}[!h]
\begin{center}
\includegraphics[width=0.95\columnwidth,angle=0,trim=0cm 0cm 0cm 0cm,clip=true]{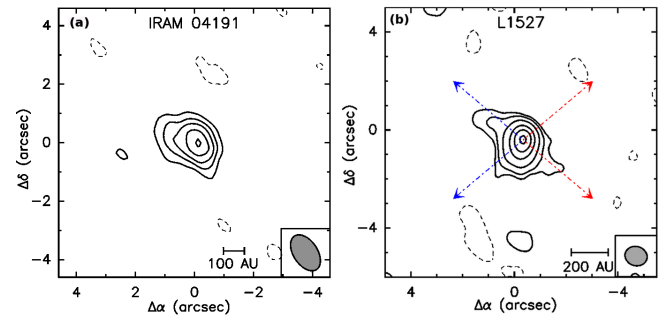}
\caption{Combined 1.3~mm dust continuum maps of the Taurus sources IRAM~04191 and L1527. 
The filled ellipses in the lower right corner of the panels indicate the synthesized HPBW beam.
{\it{(a)}} IRAM~04191. The synthetized beam (HPBW) is 1.37$\arcsec \times$ 0.82$\arcsec$, and the $1\sigma$ noise level is $\sim$ 0.31 mJy/beam. 
The level of the dashed contours is  $-3\sigma$. The solid contours correspond to levels of 3$\sigma$, 5$\sigma$, 7$\sigma$, 10$\sigma$, and 13$\sigma$.
{\it{(b)}} L1527. HPBW is 0.88$\arcsec \times$ 0.79$\arcsec$, and $\sigma$ is $\sim$ 2.0 mJy/beam. The dashed contour level is $-3\sigma$. 
The solid contours are levels of 3$\sigma$, 5$\sigma$, and 10$\sigma$ to 40$\sigma$ by 10$\sigma$. The four arrows indicate the cavity walls of the bipolar outflow driven by L1527, as traced by the emission maps of $^{12}$CO(3--2) and HCO$^{+}$(1--0) \citep{Hogerheijde98}.}
\label{fig:taurus_comb}
\end{center}
\end{figure}
%%%%%%%%

\noindent All three combined 1.3~mm continuum maps show extended emission features, 
which correlate well with the outflow cavity walls delineated by CO line observations of these Class~0 objects at similar angular resolution 
(see, e.g., \citealt{Jorgensen07b} for a $^{12}$CO(2--1) map of L1448-C with an angular resolution comparable to that of 
our combined 1.3~mm continuum map). 
This is particularly clear in the combined 1.3~mm continuum map of L1527 shown in Fig.~\ref{fig:taurus_comb}b, where the extended dust emission around the protostellar source delineates three arms of a cross (see also \citealt{ Motte01a}). The cross-like pattern seen toward L1527 coincides very well with the edges of the outflow cavity traced by CO(3--2) 
observations \citep{ Hogerheijde98, Chandler00}. 
Furthermore, if we compare our 1.3~mm continuum map (Fig.~\ref{fig:taurus_comb}b) 
with the HCO$^{+}$(1--0) interferometric map of \citet{Hogerheijde98}, we find that the cross-like morphology of the dust continuum emission coincides very closely with the features detected in HCO$^{+}$. 
A similar cross-like pattern for the dust continuum emission was also observed by \citet{Fuller95b} toward L1551-IRS5.
\\This indicates that at least some of the dust continuum emission observed in the immediate vicinity of Class~0 protostars is caused by column density enhancements due to compression in the cavity walls of their outflows.

\section{Discussion:  Constraints on the formation of multiple systems}

\subsection{Multiplicity rate of Class~0 protostars on $\sim$ 100 AU scales}

\noindent All 5 Class~0 protostars observed in the present pilot PdBI survey are single on scales between $\sim$75~AU and $\sim$1900~AU.
The only possible companion found is NGC1333-IRAS2A/SE, which is located $\sim$1900~AU away from the primary source NGC1333-IRAS2A.
We discuss below four possible explanations to the non-detection of close protobinary systems in our sample:
(1) small sample statistics, (2) selection effects in our sample, 
(3) limited mass (and mass ratio) sensitivity, 
(4) intrinsically small multiplicity fraction for Class~0 protostars on $\sim$75--1900~AU scales.
\\On the first point, even though we observed a small number of objects, the probability of drawing five single protostars from a binary fraction distribution of $\sim$32\%, corresponding to the binary fraction of Class~I YSOs in the same separation range \citep{Duchene07}, is only $\sim$14\%. 
This suggests that the Class~0 binary fraction may be lower than that of Class~I and Class~II YSOs, but only with very marginal statistical significance (1.5$\sigma$ confidence level) at the present stage. 
Interestingly, based on an extensive study of 189 Class~I sources,  \citet{Connelley08b} recently speculated that the opposite 
trend would be observed, namely that the companion star fraction should be larger at the Class~0 stage.
\\On the second point, the five sources observed in this study were selected based on a distance criterion mainly, and belong to two different star-forming regions. 
The sources in our sample have bolometric luminosities ranging from 0.1~L$_{\odot}$ to 10~L$_{\odot}$, suggesting that they span a relatively wide range of 
final stellar masses. 
Moreover, our sample includes two Class~0 objects embedded in clustered environments and belonging to wide systems (L1448-C, NGC~1333-IRAS2A),  
as well as three relatively isolated objects (IRAM~04191, L1521-F, and L1527). 
Albeit limited by its small size, our sample thus avoids the most obvious selection biases.
\\On the third point, our mass sensitivity is directly determined by the rms noise achieved in our high-resolution 1.3~mm continuum maps. 
If we assume that the envelope of any putative protostellar companion has similar density and temperature profiles to the envelope 
of the primary Class~0 object, our observations are sensitive to low circumstellar mass ratios ($q=M_{\rm{second}}/M_{\rm{main}}\sim 0.07$), except for the 
faintest two sources, IRAM~04191 and L1521-F, toward which we are only sensitive 
to M$_{\rm{second}}$/M$_{\rm{main}}\sim$ 0.6 -- 0.9. 
Assuming Class~0 systems have a distribution of circumstellar mass ratios similar to the distribution of stellar mass ratios observed toward T-Tauri binary systems 
\citep{Woitas01}, we estimate that the sensitivity of our  PdBI observations should 
allow us to detect $\geq 50$\% of the Class~0 binary systems with separations wider than $\sim$100~AU. 
\\On the fourth and final point, the multiplicity of Class~0 objects on scales $\sim$75 -- 1000~AU is not well known. 
While our sample of Class~0 protostars does not show any close multiple system (with separations $<$1900~AU), the early 
BIMA 2.7~mm continuum survey by \citet{Looney00} revealed a higher multiplicity rate in their sample at 2.7~mm. 
Among the nine Class~0 objects with separate envelopes observed by \citet{Looney00}, 
three close binary systems were found with separations $<$1000~AU, leading to a binary fraction of $\sim$33\% on scales between 100~AU and 2000~AU. 
We stress, however, that some of the protostellar companions detected by \citet{Looney00} at 2.7~mm could in fact be outflow features, like the secondary 3~mm source detected in our 3~mm continuum map of L1448-C (Fig.~\ref{fig:l1448c_comb}b). 
One striking example is the prototypical Class~0 protostar VLA~1623, for which \citet{Looney00} detected two 2.7~mm components 
separated by 1.11$\arcsec$ (i.e $\sim$150 AU), which they interpreted as a proto-binary system. 
Comparing the high-resolution 3.6~cm VLA image of \citet{Bontemps97} with the BIMA 2.7 mm image, it appears that the western 2.7 mm BIMA component may be associated with an HH-like object named HH-A by \citet{Bontemps97}, belonging to a series of HH-like cm radio continuum knots almost aligned with the outflow axis. This suggests that the BIMA 2.7~mm emission observed toward the western component (labeled ÓVLA~1623BÓ in \citealt{Looney00}) may be strongly contaminated or even dominated by free-free emission, and may not trace the presence of a bona-fide protostellar companion. 
In the light of our PdBI findings for L1448-C, careful comparison of the results with the location of the outflows from the primary protostars and detection of 
both 1~mm and 3~mm counterparts (so that a spectral index can be derived in the millimeter range) 
are needed before the protostellar nature of secondary components detected in millimeter continuum surveys can be firmly established. 
\\Combining the results of the BIMA 2.7~mm survey of \citet{Looney00} with our PdBI 1.3~mm results allows us to enlarge the sample of Class~0 sources for which the multiplicity rate between $\sim$150~AU and $\sim$1000~AU can be discussed. 
Based on the above-mentioned arguments, we do not take into account the secondary 3~mm component detected by \citet{Looney00} close to VLA~1623, as it is probably an outflow feature similar to the one observed along the outflow axis of L1448-C. 
The BIMA 2.7~mm observations probe multiplicity on scales $\simgt$100 -- 4000~AU for Taurus and Ophiuchus sources, and $\simgt$150 -- 5000~AU for Perseus sources. 
We stress that the enlarged (PdBI $+$ BIMA) sample is not homogeneous because the BIMA and PdBI surveys have differing 
sensitivities, resolutions, and observing frequencies. Nevertheless, a simple merging of the two samples allows us to draw interesting conclusions.
Among the 9 Class~0 objects mapped with BIMA, two of them are binary systems with separations less than 1000~AU: (NGC~1333-IRAS4A1 / NGC~1333-IRAS4A2) and (IRAS~16293-2422A / IRAS~16293-2422B), while no protobinary system is detected at separations less than 500~AU. 
Since the merged sample (PdBI+BIMA) has 14 target sources, this leads to estimates of 
$\sim$14\% for the binary fraction of Class~0 protostars on $\sim$150 -- 1000~AU scales , and $\simlt$7\% for the binary 
fraction of Class~0 protostars on $\sim$150 -- 550~AU scales.
Assuming that the intrinsic binary fraction of Class~0 objects in the separation range $\sim$150 -- 1000~AU is the same as 
that of Class~I YSOs, i.e. $\sim$32\% \citep{Connelley08b}, the probability of drawing 2 Class~0 binary systems in this separation range is only $\sim$9\%.
Similarly, the probability of drawing 14 single Class~0 protostars in the separation range 150--550~AU, assuming the binary fraction of Class~I YSOs, i.e. $\sim$18\% \citep{Connelley08b}, is even lower: $\sim$6\%.
Therefore, we see that combining our sample with that of  \citet{Looney00} allows us to strengthen the trend pointed out at the 
beginning of this section, namely that the Class~0 binary fraction may be lower than that of Class~I YSOs,
at least on  scales $\sim$150 -- 550~AU. Nevertheless, the trend is only present at the $\simlt 1.9\sigma $ confidence level in 
the enlarged sample, and thus remains only marginally significant. 
Clearly, more interferometric observations of Class~0 objects taken at comparable angular resolutions would  be 
needed to confirm this trend and firmly establish that binary properties evolve between the Class~0 and the Class~I stage.

\subsection{Comparison with numerical models of binary fragmentation}

%
%%%%%%%%
\begin{figure*}[!ht]
\begin{center}
\includegraphics[width=0.99\textwidth,angle=0,trim=0cm 0cm 0cm 0cm,clip=true]{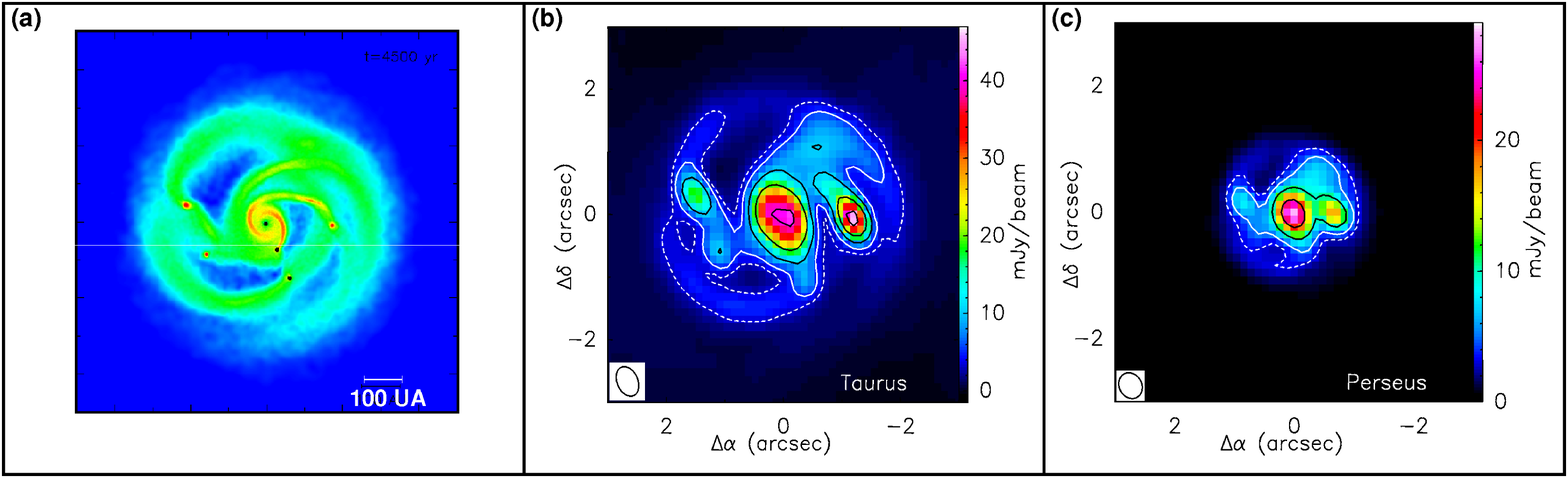}
\caption{{\it{(a)}} Model column density image from the radiative hydrodynamic simulation of Stamatellos \& Whitworth (2009) (see text for further details).
{\it{(b)}} synthetic 1.3~mm continuum image resulting from simulated A-array PdB observations of the model shown in {\it{(a)}}, assuming a distance $d=140$~pc  (distance to the Taurus complex).
{\it{(c)}} Same as (b), but assuming a distance of $d=250$~pc (distance to the Perseus complex). 
The color scale is linear and gives an indicative flux density scale in mJy/beam. The peak flux density is found to be $\sim$46~mJy/beam and $\sim$29~mJy/beam in {\it{(b)}} and {\it{(c)}}, respectively.
In both panels, the white dashed contour is showing the typical 3$\sigma$ detection level ($\sim$3 mJy/beam) achieved in our PdB-A observations, while the first plain contour is the typical 5$\sigma$ level. The following black contours are levels of 10, 20 and 40$\sigma$.
The ellipse in the bottom-left corner represents the beam size of PdB-A at the corresponding declination.}
\label{fig:Stama_1}
\end{center}
%%%%%%%%	
\vspace{1cm}
%
%%%%%%%%
\begin{center}
\includegraphics[width=0.99\textwidth,angle=0,trim=0cm 0cm 0cm 0cm,clip=true]{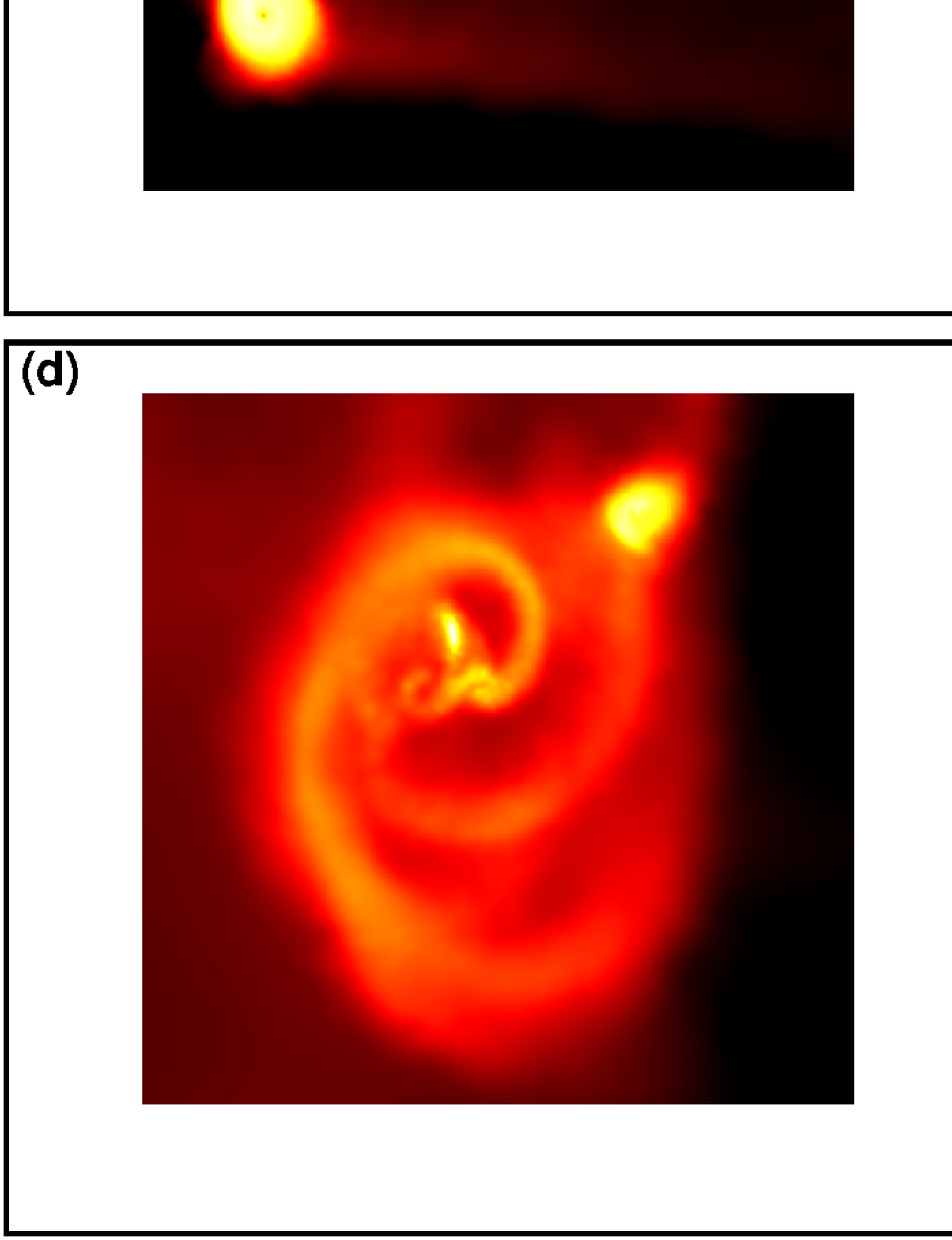}
\caption{Model column density images and synthetic 1.3~mm continuum images resulting from simulated A-array PdB observations of two typical outcomes from the model by \citet{Bate09b} (seen 2.6$\times$10$^{5}$ years after the start of the collapse, i.e 1.4 times the initial free fall time of the cloud).
In each row the first image is the model column density image, while the two images on the right are the synthetic images obtained after convolution with the PdBI A-array configuration.
(a) is a $\sim$1000~AU wide snapshot, while (d) is $\sim$600~AU wide.
(b) and (e) were produced assuming distances $d=140$~pc (similar to the distance of the Taurus complex).
(c) and (f) were produced assuming $d=250$~pc (similar to the distance of the L1448 complex). 
In all panels, the contours are levels of 3, 5, 10, 20 and 40$\sigma$, as achieved in our observations.
The color scale is linear and gives an indicative flux density scale in mJy/beam. The dashed contour is showing the typical 3$\sigma$ detection level ($\sim$3 mJy/beam) achieved in our PdB-A observations. The ellipse in the bottom-left corner represents the beam size of PdB-A at the corresponding declination.}
\label{fig:Bate_simu}
\end{center}
\end{figure*}
%%%%%%%%

In this section, we compare the results of our high-resolution PdBI observations with the predictions of three published numerical models of star formation, 
in terms of multiplicity and spatial structure. 
The first model \citep{Stamatellos09} deals with the fragmentation of a massive disk around an already formed YSO of comparable mass. 
The second hydrodynamic simulation \citep{Bate09b} describes the collapse and fragmentation of a $50\, M_\odot $ cluster-forming clump, 
and takes into account radiative feedback from formed protostellar objects. 
The third model \citep{Hennebelle08a, Hennebelle09} includes the effect of magnetic fields and simulates 
the collapse of an individual cloud core into a protostellar system.
\\In order to compare the typical outcomes of these simulations with our PdBI observations, the model column density images (in g.cm$^{-2}$)  
were put to the distance and declination of the Taurus and Perseus clouds, then converted into flux density maps (in mJy/beam) assuming optically thin dust emission at 1.3~mm, 
and $\kappa_{1.3mm} = 0.01\, {\rm cm}^2 {\rm g}^{-1}$ \citep{Ossenkopf94} and $T_d = 10 $~K for the dust properties. 
The resulting maps were convolved with the typical uv-coverage of the PdBI in A configuration to produce the synthetic 1.3~mm continuum images presented 
in Figs.~8--10 below.

\subsubsection{Numerical simulations without magnetic fields}

 {\bf{\it{~~Disk fragmentation model}}}

The hydrodynamic simulations of \citet{Stamatellos09}, performed with the SPH code {\it{DRAGON}}, 
demonstrate that the outer parts of massive extended disks are likely to undergo gravitational fragmentation, thus forming low-mass companions. 
Because they fragment rapidly, such massive disks are unlikely to be observable beyond the Class~0 phase. 
We stress that protostellar collapse is not modeled in these simulations which have no protostellar envelope component.

\noindent Figure~\ref{fig:Stama_1}(a) shows the model column density image resulting from an 
hydrodynamic simulation of a $0.7 M_{\odot}$ disk around a $0.7M_{\odot}$ star, as seen $\sim$4500 years after the start of the simulation 
(see Fig.~1 of \citealt{Stamatellos09} for more details).
Figure~\ref{fig:Stama_1}(b) and Figure~\ref{fig:Stama_1}(c)  present synthetic 1.3~mm continuum images resulting from simulations of A-array PdB observations of the model placed at the distance of Taurus and Perseus, respectively. 
One can see from the white contour and first black contour (corresponding to the average 3$\sigma$ and  5$\sigma$ levels achieved in our PdBI observations, respectively) that we expect the massive, extended disk of the model to be detected as a strong, well-resolved structure in IRAM PdBI observations. 
A circular gaussian fit to the visibilities of the synthetic images shown in Figures~8b and 8c leads to a FWHM diameter of $\sim 3.8\arcsec \pm 0.7$\arcsec at the Taurus distance, and $\sim 1.8\arcsec \pm 0.4$\arcsec at the Perseus distance: this is one order of magnitude greater than the FWHM computed in the same way for the 5 target sources detected in our PdBI maps (see Table~3), which show that all 5 target sources are compact when observed with the A array.
Furthermore, two of the Taurus sources we observed (IRAM~04191 and L1521-F) have peak fluxes which are more than one order of magnitude weaker than the peak flux in the synthetic image of Fig.~\ref{fig:Stama_1}(b). 

Taken at face value, therefore, our PdBI results are not consistent with the model of \citet{Stamatellos09}.
Note, however, that somewhat less massive ($\sim$0.1~$M_{\odot}$) disks, or initially massive disks observed at a later evolutionary stage, could be seen as compact structures, at the sensitivity achieved in our PdBI observations. 
One possible explanation for the absence of massive extended disks in our observations may be that the massive disks of the model 
are short-lived ($\sim 10^4 $~yr), as pointed out by \citet{Stamatellos09}. 
On the other hand, the presence of massive, infalling envelopes around the Class~0 objects we observed (which are not modeled in the simulations by \citealt{Stamatellos09}) should tend to refill any extended disk present at early times.
A more likely explanation in the context of this model is that only $\sim $~20--30\% of all solar-type protostars 
may develop massive extended disks at any time in their evolution \citep{Stamatellos09}. 
Our present sample is clearly not large enough to rule out this possibility. 

\noindent Therefore, while current numerical simulations of massive, extended disks do not satisfactorily reproduce our PdBI observations, 
more numerical simulations and more high-resolution observations of Class~0 objects would be needed before robust conclusions can be drawn
on this scenario.

~~\\

{\bf{\it{Hydrodynamic model including cloud collapse, disk formation, and radiative feedback}}}

In recent hydrodynamical simulations, \citet{Bate09b} treats both cloud collapse and protostar/disk formation, including the effect of radiative feedback 
from newly formed protostellar objects.

The two model snapshots presented in Fig.~\ref{fig:Bate_simu}a and Fig.~\ref{fig:Bate_simu}d (see also Fig.4 of \citealt{Bate09b}) were processed 
through the PdB simulator to produce the synthetic A-array 1.3~mm continuum images shown in Fig.~\ref{fig:Bate_simu}b,e and Fig.~\ref{fig:Bate_simu}c,f, 
with the models placed at the distances of Taurus and Perseus, respectively. 
One can see from the white level and first black level (corresponding to the average 3$\sigma$ and 5$\sigma$ levels achieved in our PdBI observations, respectively) 
that the large disk-like rotating structures produced by the models are expected 
to be detected as strong, extended or multiple sources in A-array PdBI observations. A circular gaussian fit to the visibilities of the synthetic images shown in Figures~9e and 9f leads to a FWHM diameter of $\sim 3.4\arcsec \pm 0.7$\arcsec at the Taurus distance, and $\sim 1.9\arcsec \pm 0.5$\arcsec at the Perseus distance: this is again one order of magnitude greater than the FWHM diameter measured in the same way for the 5 target sources detected in our PdBI maps (see Table~3), which show that all 5 target sources are compact when observed with the A array.

~~\\
We conclude that purely hydrodynamic simulations, even if they include radiative feedback which inhibits fragmentation close to existing protostellar objects, 
fail to reproduce our PdBI observations, because they tend to form massive extended structures (with FWHM $\sim$ 300--500~AU) and/or multiple systems. 
One should bear in mind, however, that the absolute age  of Class~0 objects ($\sim 3\times 10^4$--$2\times 10^5$~yr) is quite uncertain (cf. \citealt{Evans09}), 
which casts some doubt on the time at which the models should be compared to our observations.

\subsubsection{Magnetohydrodynamic model}

%
%%%%%%%%
\begin{figure*}[!ht]
\begin{center}
\includegraphics[width=0.85\textwidth,angle=0,trim=0cm 0cm 0cm 0cm,clip=true]{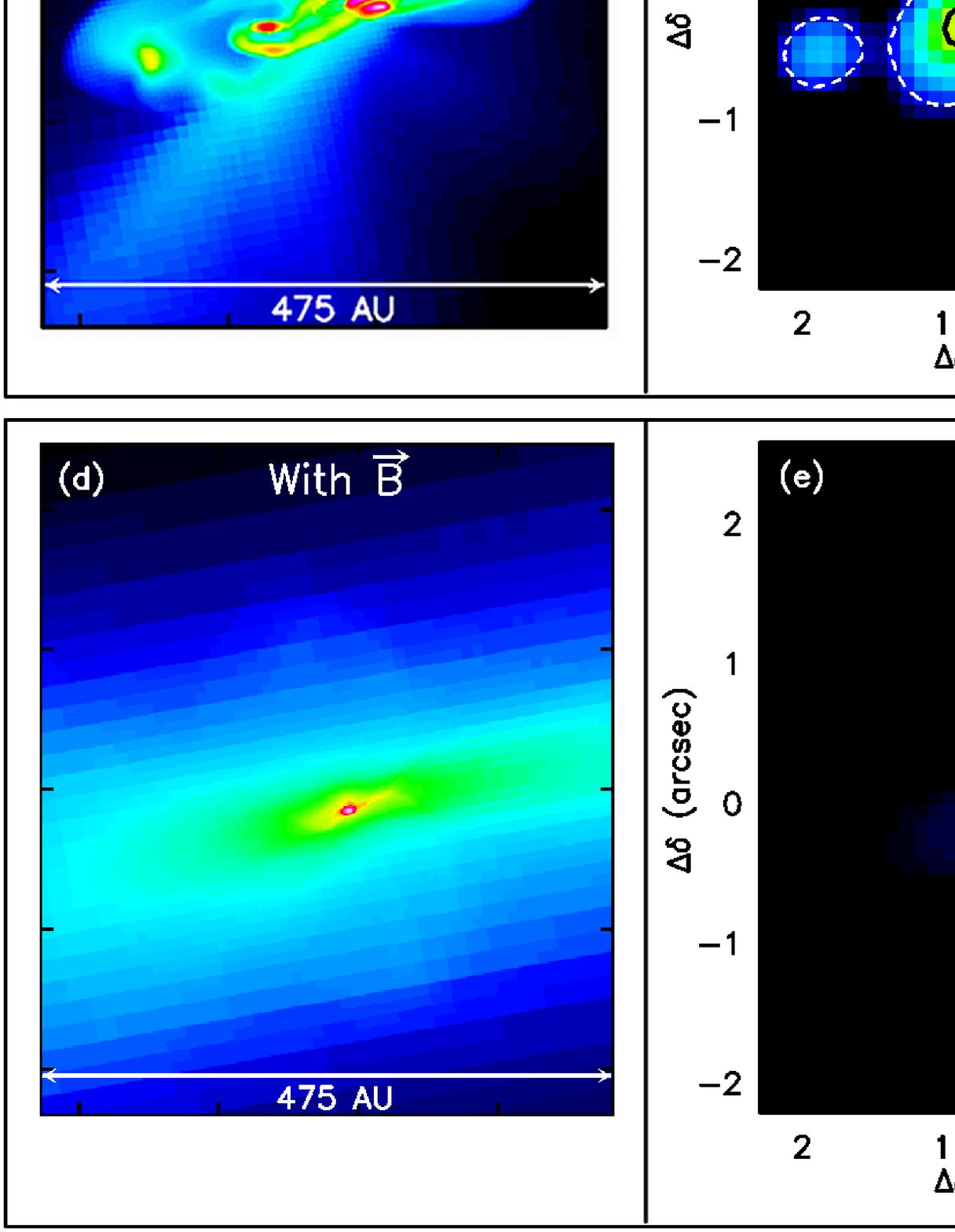}
\caption{Synthetic 1.3~mm continuum images resulting from simulated A-array PdB observations of two typical outcomes from the simulations of magnetized core collapse from e.g. \citet{Hennebelle08b} (see text for further details).
{\it{Upper row}}: Panel (a) shows a model snapshot view of the column density distribution of the inner part of a  
protostellar system obtained $\sim 10^4$~yr after the beginning of collapse in purely hydrodynamic ($B = 0$) simulations of cloud core collapse. Panel (b) shows the synthetic PdB-A 1.3~mm dust continuum image produced from the purely hydrodynamic model (a), put to the distance of the Taurus complex ($d=140$~pc). Panel (c) shows a synthetic image similar to (b), but put to the distance of the Perseus complex ($d=250$~pc).
{\it{Lower row}}: Panel (d) shows a snapshot view of the column density distribution of the inner part of a  
protostellar system obtained from MHD simulations starting from the same initial conditions as in panel (a), except for a non-zero magnetic field (whose initial value is $1/2$ of the critical field strength required to prevent collapse).  Panel (e) shows the synthetic PdB-A 1.3~mm dust continuum image produced from the MHD model shown in (d), and put to the distance of the Taurus complex ($d=140$~pc). Panel (f) shows a synthetic image similar to (e), but put to the distance of the Perseus complex ($d=250$~pc).
In all of the synthetic images, the color scale is linear and gives an indicative flux density scale in mJy/beam, assuming  $T_d = 10 $~K and $\kappa_{1.3mm} = 0.01\, {\rm cm}^2 {\rm g}^{-1}$ for the dust properties. The white dashed contour shows the typical 3$\sigma$ detection level ($\sim$3 mJy/beam) achieved in our PdB-A observations, while the following black contours are: 5$\sigma$ (thick), 10$\sigma$, 20$\sigma$ and 40$\sigma$.}
\label{fig:Hennebelle_simu}
\end{center}
\end{figure*}
%%%%%%%%

Adaptive Mesh Refinement (AMR) simulations of cloud core collapse and fragmentation were carried out by \citet{Hennebelle08a} and \citet{Hennebelle08b}, using the MHD version of the {\it{RAMSES}} code (see \citet{Fromang06}.
In these MHD simulations, the formation of a centrifugally-supported disk is suppressed by magnetic braking  (see \citealt{Mellon09}). 
Moreover, \citet{Hennebelle08b} showed that, for rotating dense cores with a magnetic field strength 
typical of values inferred from observations (e.g. Crutcher 1999),  
fragmentation was suppressed by magnetic fields if the initial density perturbations were too small.   
\\Two typical outcomes of these simulations are shown in Figure~\ref{fig:Hennebelle_simu}: 
Fig.~\ref{fig:Hennebelle_simu}a and Fig.~\ref{fig:Hennebelle_simu}d are model column density images obtained in the absence and in the presence of a 
magnetic field, respectively. Fig.~\ref{fig:Hennebelle_simu}b and Fig.~\ref{fig:Hennebelle_simu}e represent 
synthetic images obtained from these models after convolution with the dirty beam of PdBI in A configuration and cleaning, assuming a distance $d=140$~pc, while Fig.~\ref{fig:Hennebelle_simu}c and Fig.~\ref{fig:Hennebelle_simu}f represent the same synthetic images, but assuming a distance $d=250$~pc.
The simulations shown in the upper and lower panels started from a rotating, centrally-condensed spherical core 
0.013~pc in radius, with a density contrast of 10. The total mass of the core was $2 M_{\odot}$, 
and the two simulations started from the same initial conditions, 
except for the initial value of the magnetic field ($B = 0$ in the first case; moderate B-field strength in the second case, i.e., initial mass-to-flux ratio $\mu=3.2$
in units of the critical value for collapse).
Note that the purely hydrodynamic simulation of Fig.~\ref{fig:Hennebelle_simu}a
produces a quadruple system within a radius of $\sim 200$~AU, 
while core fragmentation is completely suppressed in the moderately magnetized simulation of Fig.~\ref{fig:Hennebelle_simu}d, which 
leads to the formation of a single stellar object within 200~AU.
\\The magnetized model shown in the lower panels of Fig.~\ref{fig:Hennebelle_simu} reproduces our PdBI observations 
of the Taurus Class~0 protostars quite well: the synthetic image (Fig.~\ref{fig:Hennebelle_simu}e) 
shows a unique source, with a peak flux density $\sim$3--8 mJy/beam, and a size comparable to those observed : a circular gaussian fit to the visibilities of the synthetic images shown in Figures~10e leads to a FWHM diameter of $\sim$0.4\arcsec ~at the Taurus distance ($\sim$0.2\arcsec ~at the Perseus distance), comparable to the ones computed for the sources detected in our PdBI maps, and given in Table~3.
This comparison suggests that magnetic fields are an essential ingredient of the early phases of star formation, 
as they seem to be a plausible way to regulate core/disk fragmentation 
and obtain single objects on scales $<$ 300 AU, similar to what is observed in our sample.

\subsubsection{Implications for the formation of multiple systems}

Although our sample clearly needs to be extended before general conclusions can be drawn, the non-detection of 
multiple Class~0 systems with separations between $\sim$~100~AU and $\simlt$~600~AU in both our sample and the BIMA sample 
of \citet{Looney00} already suggests interesting implications for the formation of  solar-type multiple systems.
\\Two alternative scenarios can be proposed. 
First, {\bf{it}} is possible that wide ($> 600$~AU) multiple systems form during the Class~0 phase, 
and that orbital migration then reduces the separation between the protostellar components during evolution to the Class~I phase \citep{Bate00a, Bate02a}. 
Observations \citep[see, for example][]{Connelley08b} suggest that main-sequence systems are tighter than T Tauri systems, and that the latter themselves have 
separations which are somewhat smaller than those observed toward 
Class~I YSOs \citep{Patience02}. 
Therefore, it is possible that the typical separation of multiple systems decreases in the course of YSO evolution \citep{Connelley08b}, even if such an effect is not yet well understood. 
The formation of wide binary systems at the Class~0 stage has been envisaged by \citet{Price07} and \citet{Hennebelle08b}. 
Based on their MHD numerical simulations, these authors showed that, in the presence of large-amplitude initial perturbations, 
each perturbation develops independently leading to the formation of a wide protobinary system, which can then gravitationally contract to typical PMS binary separations.\\
A second possible scenario would be that multiple systems form with tight separations $\leq$75~AU, explaining 
the paucity of systems in the $\sim $~100--500~AU separation range at the Class~0 stage, but that these 
systems then expand to produce multiple systems with wider typical separations at the Class~I stage. 
This alternative scenario is plausible if fragmentation occurs during the second collapse phase, 
a possibility which has been explored in the numerical simulations of \citet{Bonnell94} and \citet{Machida07}.
One of the conditions for this model to produce binary systems with typical separations $\sim$ 100 -- 500 AU is that the initially very tight 
protobinary system must gain sufficient angular momentum by accretion (see e.g. \citealt{Goodwin04, Bate00a}) 
to increase the separation between the two protostellar components by the end of the Class~0 phase.

\section{Summary and conclusions}

We carried out a subarcsecond pilot study of 5 Class~0 objects at 1.3~mm with the IRAM Plateau de Bure Interferometer in its most extended configuration, 
in an effort to probe protostellar multiplicity at separations $50<a<5000$~AU at the beginning of the embedded YSO phase.
Continuum emission and $^{12}$CO(2--1) emission were observed simultaneously, with a typical resolution $\sim 0.3\arcsec $--0.5\arcsec 
and rms sensitivity $\sim$~0.1--1 mJy/beam, which allowed us to study multiplicity down to separations $a \sim$50 AU and 
circumstellar mass ratios $q \sim$0.07.\\

\noindent Our main results and conclusions can be summarized as follows:

\begin{enumerate}
\item All five primary Class~0 protostars (IRAM~04191, L1527, L1521-F, L1448-C, and NGC1333-IR2A) were detected in the 1.3~mm continuum maps, with signal-to-noise ratios ranging from 5.4 to 65.5. 
\item Toward L1448-C, a series of seven high-velocity ($v_{LSR}>$ 50~km s$^{-1}$) bullets were detected in $^{12}$CO(2--1), which trace the axis of the bipolar jet in both the redshifted and blueshifted lobes of the outflow.
\item Single 1.3~mm continuum components associated with the primary Class 0 objects were detected in the case of the three Taurus sources, while robust 
evidence of secondary components was found toward the two Perseus sources, L1448-C and NGC1333-IR2A.
\item The L1448-C secondary component lies $\sim$ 600~AU south-east of the primary source, at a position angle close to that of the CO(2--1) jet axis. We show that it is not a genuine protostellar companion but rather an outflow feature directly associated with the powerful jet driven by L1448-C. 
The detection of compact millimeter continuum emission originating from such an outflow-generated feature emphasizes the need to observe outflow-shock diagnostics, before any robust conclusion can be drawn on the nature of secondary components detected in the vicinity of protostellar objects.
\item The nature of the NGC1333-IR2A secondary component, detected $\sim$ 1900~AU south-east of the primary source, is as yet unclear: it may either be a genuine protostellar companion or trace the edge of an outflow cavity.
\item Altogether, our pilot PdBI survey found only evidence of outflow-generated features, and/or wide protobinary systems: no multiple system was detected at separations $a<$1900~AU in our sample of 5 Class~0 protostars.
\item Combining our results with the BIMA survey of 9 Class~0 objects by \citet{Looney00}, we argue that there is no evidence of multiple protostellar systems in the separation range $150<a<550$~AU among an enlarged sample of 14 Class~0 protostars. 
Although the millimeter interferometric observations available for this enlarged sample are inhomogeneous, they tentatively suggest that the Class~0 binary fraction may be lower than that of Class~I YSOs, at least for separations $\sim$150 -- 500 AU. This tentative evolution of the binary fraction from the Class~0 to the Class~I stage is present at the $\sim$1.9$\sigma$ confidence level in the enlarged sample, and thus requires confirmation.
\item Comparison of synthetic model images with our PdBI results shows that purely hydrodynamic models of protostellar collapse and disk formation have difficulties matching our observations, since these models typically produce multiple components, embedded in large-scale rotating structures, which are not observed toward our sample of five Class~0 sources. These large-scale rotating structures may be short-lived, however, and 
more observations would be needed to draw robust conclusions, given the currently large uncertainties on the Class~0 lifetime.
\item Comparison of synthetic model images from magnetohydrodynamic models with our PdBI results shows that magnetized models of protostar formation 
agree better with our observations, as magnetic fields tend to prevent the formation of extended disk-like structures and to suppress fragmentation into multiple components on small scales (100~AU -- 1000 AU). 
\item However, magnetohydrodynamic models may allow wide ($\geq$ 1000 AU) and/or very tight ($\leq$ 30 AU) multiple systems to form during the Class~0 phase. The paucity of multiple Class 0 systems with separations 150~AU$\leq a \leq$ 600 AU, if confirmed by comparable observations of larger source samples, may thus favor binary formation scenarios which involve dynamical evolution of the system separations with time. 
\end{enumerate}
~
\newline {\it{Acknowledgements}}~   
We are grateful to the IRAM-PdBI staff, and more specifically to Roberto Neri and Fr\'ed\'eric Gueth, for their precious help with the PdBI observations and data processing.
The work presented in this paper was stimulated by discussions held in the context of the Marie Curie Research Training Network ``Constellation'' (MRTN-CT2006- 035890). 

\clearpage

    \bibliographystyle{aa}
      \bibliography{maury_13492_text}

\end{document}